\documentclass[prl,longbibliography,twocolumn,nopacs,preprintnumbers,notitlepage,amsmath,amssymb,superscriptaddress]{revtex4-2}

\usepackage{amsmath}
\usepackage{amssymb}
\usepackage{amsbsy}
\usepackage{graphicx}
\usepackage{xcolor}

\usepackage{mathrsfs}

\usepackage{enumerate}
\usepackage{mathtools}

\usepackage{soul}

\newcommand{\moy}[1]{\left\langle #1 \right\rangle}
\newcommand{\dd}[0]{\mathrm{d}}

\newcommand{\moyE}[1]{\left\langle #1 \right\rangle}
\newcommand{\moyI}[1]{\overline{#1}}

\newcommand{\psiA}[0]{\psi_{\mathrm{A}}}
\newcommand{\psiQ}[0]{\psi_{\mathrm{Q}}}

\newcommand{\wA}[0]{w_{\mathrm{A}}}
\newcommand{\wQ}[0]{w_{\mathrm{Q}}}

\newcommand{\Gt}[2]{\tilde{G}_{t}^{(\lambda)}(#1 | #2)}

\def\e{e}
\def\I{\mathrm{i}}

\def\Ot{\mathcal{O}_t}

\newcommand{\sg}[1]{\ensuremath{\mathrm{sign}\left( #1 \right)}}
\newcommand{\abs}[1]{\ensuremath{\left| #1 \right|}}

\DeclareMathOperator{\erfc}{erfc}

\definecolor{darkblue}{rgb}{0,0,0.6}
\definecolor{darkred}{rgb}{0.6,0,0}

\usepackage[colorlinks=true,urlcolor=darkblue,citecolor=darkblue,linkcolor=darkred]{hyperref}

\begin{document}

\title{From Particle Currents to Tracer Diffusion: Universal Correlation Profiles in Single-File Dynamics}

\author{Aur\'elien Grabsch}
\affiliation{Sorbonne Universit\'e, CNRS, Laboratoire de Physique Th\'eorique de la Mati\`ere Condens\'ee (LPTMC), 4 Place Jussieu, 75005 Paris, France}

\author{Th\'eotim Berlioz}
\affiliation{Sorbonne Universit\'e, CNRS, Laboratoire de Physique Th\'eorique de la Mati\`ere Condens\'ee (LPTMC), 4 Place Jussieu, 75005 Paris, France}

\author{Pierre Rizkallah}
\affiliation{Sorbonne Universit\'e, CNRS, Physico-Chimie des \'Electrolytes et Nanosyst\`emes Interfaciaux (PHENIX), 4 Place Jussieu, 75005 Paris, France}

\author{Pierre Illien}
\affiliation{Sorbonne Universit\'e, CNRS, Physico-Chimie des \'Electrolytes et Nanosyst\`emes Interfaciaux (PHENIX), 4 Place Jussieu, 75005 Paris, France}

\author{Olivier B\'enichou}
\affiliation{Sorbonne Universit\'e, CNRS, Laboratoire de Physique Th\'eorique de la Mati\`ere Condens\'ee (LPTMC), 4 Place Jussieu, 75005 Paris, France}

\begin{abstract}

Single-file transport refers to the motion of particles in a narrow channel, such that they cannot bypass each other. This constraint leads to strong correlations between the particles, described by correlation profiles, which measure the correlation between a generic observable and the density of particles at a given position and time. They have recently been shown to play a central role in single-file systems.
Up to now, these correlations have only been determined for diffusive systems in the hydrodynamic limit. Here, we consider a model of reflecting point particles on the infinite line, with a general individual stochastic dynamics. We show that the correlation profiles take a simple universal form, at arbitrary time.
We illustrate our approach by the study of the integrated current of particles through the origin, and apply our results to representative models such as Brownian particles, run-and-tumble particles and Lévy flights.
We further emphasise the generality of our results by showing that they also apply beyond the 1d case, and to other observables.

\end{abstract}

\maketitle

\let\oldaddcontentsline\addcontentsline
\renewcommand{\addcontentsline}[3]{}

\emph{Introduction.---} Single-file transport, where particles move in narrow channels with the constraint that they cannot bypass each other, has become a fundamental model for transport in confined systems~\cite{Levitt:1973,Arratia:1983,Hahn:1996,Wei:2000,Lin:2005}.
Experimentally, this situation has been observed in various physical, chemical or biological systems, such as zeolites, colloidal suspensions, or carbon nanotubes~\cite{Hahn:1996,Wei:2000,Lin:2005,Cambre:2010}. Theoretically, it is a central field of statistical physics, relevant both at equilibrium and out-of-equilibrium~\cite{Krapivsky:2010,Chou:2011}.
In this context, two key observables have received a notable attention: (i) the integrated current through the origin $Q_t$ (defined as the number of particles which have crossed the origin from left to right, minus those from right to left, up to time $t$)~\cite{Derrida:2009,Derrida:2009a,Krapivsky:2012,Banerjee:2020,Mallick:2022,Bettelheim:2022,Dean:2023}; (ii) the position $X_t$ of a tracer~\cite{Levitt:1973,Harris:1965,Arratia:1983,Ryabov:2012,Krapivsky:2014,Krapivsky:2015,Sadhu:2015,Cividini:2016,Cividini:2016a,Kundu:2016,Poncet:2021,Grabsch:2022,Grabsch:2023}, which can be monitored experimentally at various scales~\cite{Hahn:1996,Wei:2000,Lin:2005}.

Because the order of the particles is conserved at all times, strong correlations between these observables and the density of particles $\rho(x,t)$ emerge. For instance, an increase of $Q_t$ imposes a density at the right of the origin higher than average, and a lower on the left. A similar effect occurs with $X_t$: a large displacement of the tracer in a given direction involves the displacement of more and more particles in the same direction. This leads to a striking subdiffusive behaviour $\langle X_t^2\rangle \propto\sqrt{t}$~\cite{Harris:1965} in contrast with the regular diffusion $\langle X_t^2\rangle \propto{t}$.

Despite their importance, the quantification of the coupling between $Q_t$ or $X_t$ and $\rho(x,t)$ remains a broadly open question. Recently, they have been characterised for the Symmetric Exclusion Process, and other paradigmatic models of single-file diffusion~\cite{Poncet:2021,Grabsch:2022,Grabsch:2023}.
In addition to their clear physical relevance, these correlations have also acquired a technical importance since they have been shown to satisfy a closed equation for these systems~\cite{Poncet:2021,Grabsch:2022,Grabsch:2023}\footnote{Note that, since the publication of \cite{Grabsch:2022} in which this closed equation first appears, the exactness of the equation has been proved \cite{Mallick:2022} using the inverse scattering method (see also \cite{Bettelheim:2022,Krajenbrink:2023}).}. 
However, these results are limited to (i) the case of diffusive systems (in which the individual particles have a diffusive motion); (ii) the long time behaviour; (iii) the  specific case of $X_t$ and $Q_t$.

Here, by considering a model of reflecting point particles on the infinite line, with an arbitrary individual stochastic dynamics, we overcome these limitations. We show that the correlation profiles take a simple universal form (with respect to the individual motion of the particles), at arbitrary time, and for a large class of observables (as defined below).

More precisely, we illustrate our approach by the study the integrated current of particles through the origin, and apply our results to representative processes which go beyond Brownian particles, such as (i) run-and-tumble particles which are a key model to describe active transport~\cite{Weiss:2002,Tailleur:2008}; and (ii) Lévy flights which is an emblematic model of superdiffusion~\cite{Bouchaud:1990}.
We further emphasise the generality of our results by showing that they also apply beyond the 1d case, and to other observables.

\begin{figure}
    \centering
    \includegraphics[width=\columnwidth]{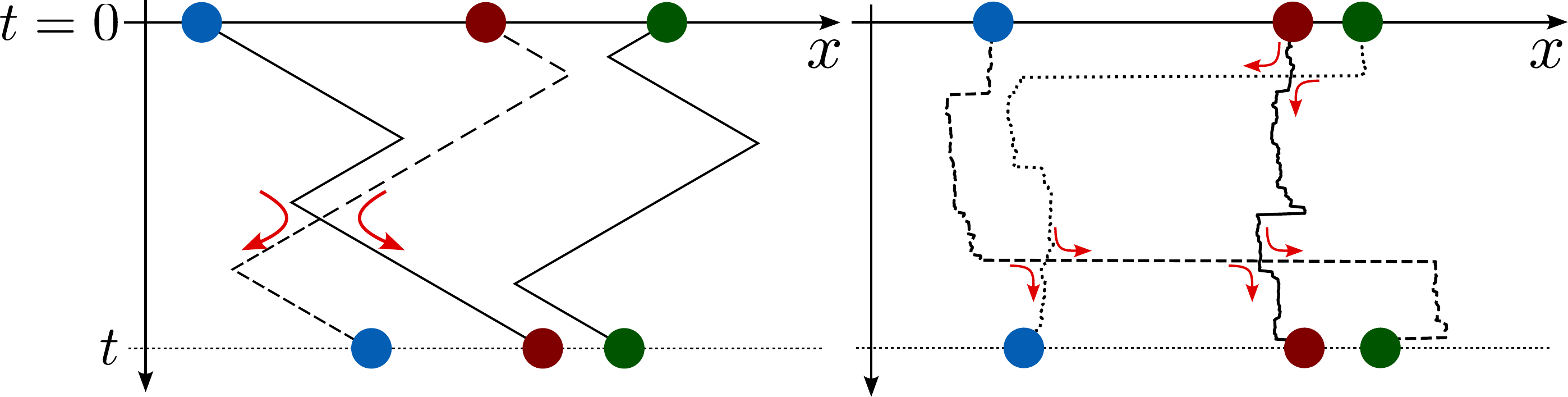}
    \caption{The motion of reflective particles (illustrated by the arrows) can be mapped onto the motion of noninteracting particles (solid and dashed lines). Left: illustration with by run-and-tumble particles, which move at a constant speed, and flip their direction at random times. When two particles collide, they flip their direction. Right: Lévy flights. When a particle collides another, it stops and the next particle is pushed. This can lead to a series of collisions.}
    \label{fig:SchemaRTP}
\end{figure}

\emph{Model.---} We first consider $N$ particles on the real line, with position $\{ x_i(t) \}_{i=1,\ldots,N}$ at time $t$. In a second step, we will take the thermodynamic limit $N \to \infty$. Initially, the $N$ particles are independently picked from a density $\rho_0(x)$, normalised such that $\int \rho_0(x) \dd x = N$. Each particle has a stochastic dynamics in time, described by its propagator $G_t(x|y)$, i.e. the prob. to find the particle at position $x$ at time $t$, knowing that it was at position $y$ at time $0$.
When this dynamics leads to a crossing between two particles, they are simply exchanged, leading to a reflection of the particles. This dynamics can be mapped onto the one of \textit{noninteracting} particles (see Fig~\ref{fig:SchemaRTP}).  While this formally applies to any propagator, this is especially relevant for Markovian dynamics, since the definition of the contact is more tricky in the non-Markovian case.
The time evolution of the particles being independent from their initial distribution, we define two different types of averaging: (i) the average over the time evolution of the particles, denoted for any function $f$,
\begin{equation}
    \label{eq:DefMoyE}
    \moyE{f(\{ x_i(t) \})} = 
    \int \prod_{i=1}^N \dd x_n \: K_t(\{ x_i \} | \{ x_i(0) \}) \: f(\{ x_i \})
    \:,
\end{equation}
with $K_t$ the $N$-particles propagator, and (ii) the average over the initial positions,
\begin{equation}
    \label{eq:DefMoyI}
    \moyI{f(\{ x_i(0) \})} = 
    \int \prod_{n=1}^N \dd y_n \frac{\rho_0(y_n)}{N} f(\{ y_i \})
    \:.
\end{equation}
Although our approach can be applied to many observables, we will focus for concreteness on the integrated current through the origin, which measures the variation of the number of particles on the positive axis,
\begin{equation}
    \label{eq:DefCurrentQt}
  Q_t = \sum_{i} \left[ \Theta(x_i(t)) - \Theta(x_i(0)) \right]
  \:,
\end{equation}
where $\Theta$ is the Heaviside step function. We are interested in the statistical properties of this observable, and its correlations with the global density of particles
\begin{equation}
  \rho(x,t) = \sum_i \delta(x - x_i(t))
  \:.
\end{equation}
These two quantities are indeed expected to be strongly correlated. These correlations are encoded in the joint cumulant generating function $\ln \moyI{\moyE{\e^{\lambda Q_t + \chi \rho(x,t)}}}$, where $\lambda$ and $\chi$ are the parameters of the generating function. We have used here the \textit{annealed} averaging, as usually defined in statistical mechanics, which corresponds to averaging over both the time evolution and all the initial positions. For simplicity, we will focus on the lowest orders in $\chi$, which are the cumulant generating function of the integrated current,
$
    \psiA(\lambda,t) \equiv
    \lim_{N \to \infty} \ln \moyI{\moyE{\e^{\lambda Q_t}}}
$,
and the correlation profile~\cite{Poncet:2021}
\begin{equation}
    \wA(x,\lambda,t) \equiv
    \lim_{N \to \infty}
    \frac{\moyI{\moyE{\rho(x,t) \e^{\lambda Q_t}}}}{ \moyI{\moyE{\e^{\lambda Q_t}}}}
    \:.
\end{equation}
These correlation profiles have been show to play an important role, since they verify simple closed equations for several important models of single-file systems~\cite{Poncet:2021,Grabsch:2022,Grabsch:2023}.

We also consider the case of a \textit{quenched} initial condition, which corresponds to averaging over the time evolution of the typical initial positions of the particles. The initial condition is well-known to play a key role in single-file systems, as exemplified by "everlasting" effects on various observables~\cite{Leibovich:2013,Krapivsky:2015,Sadhu:2015,Poncet:2021a,Krug:1997}.
In this case, the joint cumulant generating function is $\moyI{\ln \moyE{\e^{\lambda Q_t + \chi \rho(x,t)}}}$. At lowest orders in $\chi$, this gives the quenched cumulant generating function
$
    \psiQ(\lambda,t) \equiv
    \lim_{N \to \infty} \moyI{\ln \moyE{\e^{\lambda Q_t}}}
$,
and the quenched correlation profile
\begin{equation}
    \wQ(x,\lambda,t) \equiv
    \lim_{N \to \infty}
    \moyI{\frac{\moyE{\rho(x,t) \e^{\lambda Q_t}}}{\moyE{\e^{\lambda Q_t}}} }
    \:.
\end{equation}

\emph{Results.---} The key ingredient is the joint propagator of the $N$ particles, which takes the form~\cite{Rodenbeck:1998}
\begin{equation}
  \label{eq:PropNpart}
  K_{t}(\vec{x} | \vec{y} ) =
  \frac{1}{N!} \sum_{\sigma} \prod_{i=1}^N G_t(x_i|y_{\sigma(i)})
  \:,
\end{equation}
where the sum runs over all permutations $\sigma$ of the $N$ particles.
Computing first the averages~(\ref{eq:DefMoyE},\ref{eq:DefMoyI}), and then taking the thermodynamic limit $N \to \infty$~\footnote{We have introduced the initial density of particles $\rho_0$ with the constraint that it is normalised to $N$. In the thermodynamic limit $N \to \infty$, this constraint relaxes and one can consider non-normalised initial densities, such as $\rho_0(x) = \rho$ constant.}, we obtain that the cumulant generating function and the correlation profiles take a simple universal form (see Supplementary Material (SM)~\cite{SM} for details of the derivation). In the annealed case,
\begin{equation}
  \label{eq:PsiAnnealedGenSF}
    \psiA(\lambda,t)
    =  \int \dd y \: \rho_0(y) \int (\Gt{x}{y}-G_t(x|y)) \dd x
    \:,
\end{equation}
\begin{equation}
  \label{eq:ProfAnnealedGenSF}
    \wA(x,\lambda,t)
    = \int \rho_0(y) \Gt{x}{y} \dd y 
    \:,
\end{equation}
where we have defined the tilted propagator
\begin{equation}
    \label{eq:DeforProp}
    \Gt{x}{y} = \e^{\lambda \Theta(x)} G_t(x|y) \e^{- \lambda \Theta(y)}
    \:.
\end{equation}
Similarly, in the quenched case,
\begin{equation}
  \label{eq:PsiQuenchGenSF}
    \psiQ(\lambda,t)
    = \int \dd y \:
    \rho_0(y) \ln \left[ \int_{-\infty}^\infty \Gt{x}{y}  \dd x \right]
    \:,
\end{equation}
\begin{equation}
  \label{eq:ProfQuenchGenSF}
   \wQ(x,\lambda,t)
    = \int \dd y \: \rho_0(y)
    \frac{\Gt{x}{y}}
    {\int_{-\infty}^\infty  \Gt{z}{y} \dd z}
    \:.
\end{equation}
These expressions hold for any propagator $G_t$ of an individual particle, for any initial density of particles $\rho_0$, and at arbitrary time $t$. They constitute the main results of this article.
Note that our results hold in presence of external forces (a situation studied for instance in~\cite{Antonov:2022})~\footnote{ Note that our results hold in the presence of external forces, but not for arbitrary interparticle interactions, since this latter case involves the full many-body propagator which cannot be reduced to a single-particle propagator in the general case.}. The key point of our derivation is that all the particles have the same dynamics (with the requirement that the memory of the past is lost upon collision), and feel the presence of the other particles only when a collision occurs. The only ingredient needed is the one particle propagator $G_t$, either analytically or numerically~\footnote{If the propagator is known analytically, our main results~(\ref{eq:PsiAnnealedGenSF}-\ref{eq:ProfQuenchGenSF}) yield explicit expressions for the cumulants and profiles. If the propagator is obtained numerically, our results provide a straightforward way to compute these observables numerically by at most two numerical integrations.}.
We now give the example of run-and-tumble particles, for which the propagator is known explicitly. This will allow us to discuss on a concrete example the physics of these correlation profiles. Formulas for the cases of Brownian particles and Lévy flights are given in SM~\cite{SM}.

\begin{figure}
    \centering
    \includegraphics[width=\columnwidth]{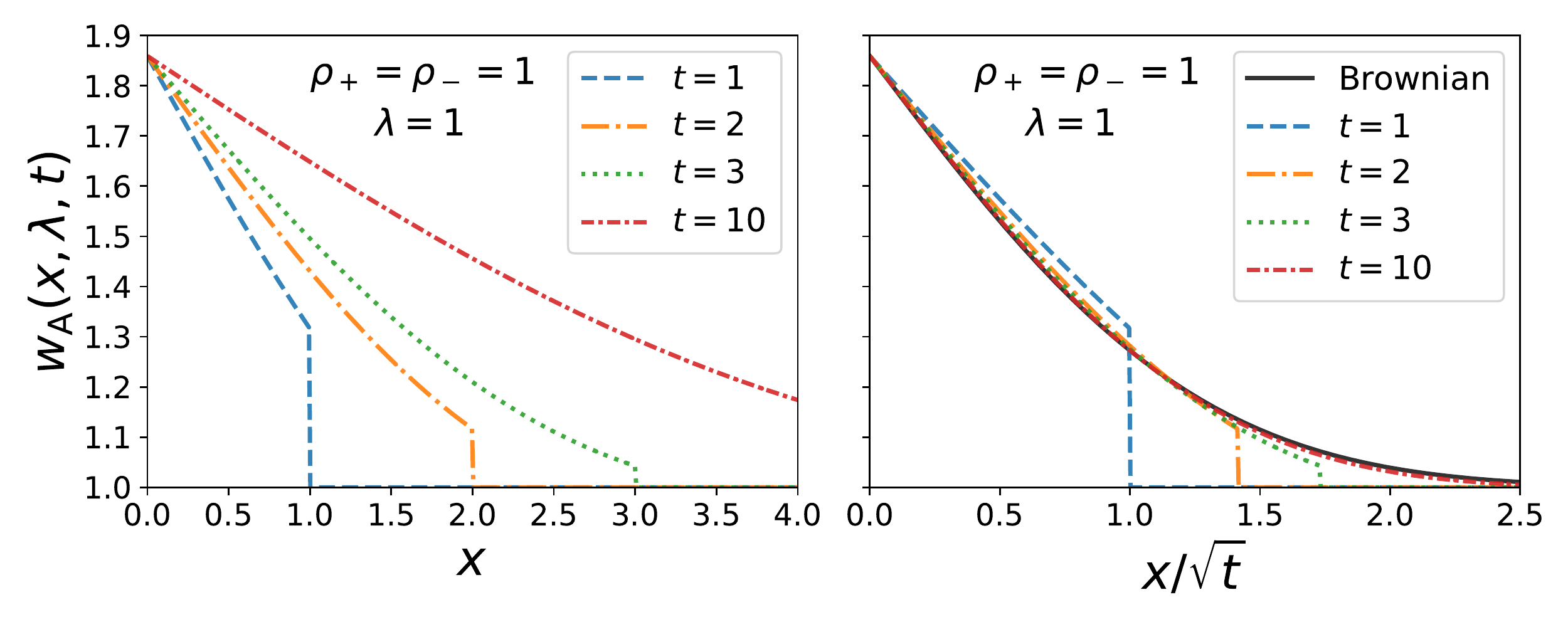}
    \caption{Annealed correlation profile $\wA$~\eqref{eq:wARTP} for run-and-tumble particles, with $v_0 = 1$ and $\gamma = 1$. Left: Profile as a function of $x$ for different times. Right: Profile as a function of the rescaled variable $x/\sqrt{t}$ at different times. For $t \to \infty$, it converges to the profile for diffusive particles (solid black lines), with a diffusion constant $D=v_0^2/(2\gamma) = \frac{1}{2}$.}
    \label{fig:ProfRTP}
\end{figure}

\emph{Application: run-and-tumble particles.---} We consider a system of run-and-tumble particles, which is an important model of active particles, involved in various contexts~\cite{Weiss:2002,Tailleur:2008}. These particles move at constant speed $v_0$, and flip their direction of motion with rate $\gamma$.  When two particles collide, they are reflected (see Fig.~\ref{fig:SchemaRTP}). For simplicity, we will consider a step initial density of particles $\rho_0(x) = \rho_+ \Theta(x) + \rho_- \Theta(-x)$. The Laplace transform of the propagator of an individual particle takes a simple form~\cite{Weiss:2002}. We can easily obtain the annealed profile and cumulant generating function in the Laplace domain since the expressions~(\ref{eq:PsiAnnealedGenSF},\ref{eq:ProfAnnealedGenSF}) are linear in the propagator. The inverse Laplace transform can be computed explicitly using the expressions given in~\cite{Banerjee:2020}, and we get
$\psiA(\lambda,t) = \frac{\omega}{2} v_0 t \: \e^{-\gamma t}
(\mathrm{I}_0(\gamma t) + \mathrm{I}_1(\gamma t))$, where we have denoted $\omega = \rho_+ (\e^{-\lambda}-1) + \rho_- (\e^{\lambda}-1)$, by analogy with the single parameter identified in the simple exclusion process (SEP)~\cite{Derrida:2009}, and $\mathrm{I}_\nu$ is a modified Bessel function. Similarly, the correlation profile reads
\begin{multline}
    \label{eq:wARTP}
    \wA(x > 0,\lambda,t) =
    \rho_+
  + \frac{\rho_- \e^\lambda - \rho_+}{2}
  \Theta(v_0 t - x)
  \\
  \times \left(
    \e^{-\frac{\gamma x}{v_0}}
    + \frac{\gamma x}{v_0}
    \int_1^{\frac{v_0 t}{x}} \frac{\e^{-\frac{\gamma x T}{v_0}}
      \mathrm{I}_1(\frac{\gamma x}{v_0} \sqrt{T^2-1})}{\sqrt{T^2-1}}
    \dd T
  \right)
  \:.
\end{multline}
This profile, represented in Fig.~\ref{fig:ProfRTP}, quantifies the correlation between the observable $Q_t$ and the density of particles $\rho(x,t)$. When $\lambda > 0$, $\wA(x>0,\lambda,t) \geq \rho_+$, indicating that an increase of the current yields an increase of the density at the right of the origin.
We emphasise that (i) our approach captures the full dynamics of the profile $\wA$~\eqref{eq:wARTP}, illustrated in Fig.~\ref{fig:ProfRTP}. In particular it presents a sharp cutoff at $x = v_0 t$, which is a consequence of the finite speed $v_0$ of the particles, showing that $Q_t$ and $\rho(x,t)$ are decorrelated for $x > v_0 t$; (ii) the dependence of the profile~\eqref{eq:wARTP} in $\rho_+$, $\rho_-$ and $\lambda$ is in fact a general feature which holds for any propagator $G_t$, and is only a consequence of the choice of the observable $Q_t$ and the initial step of density $\rho_0(x)$.

We now demonstrate that our approach can be extended in several important directions (see SM~\cite{SM} for details).

\begin{figure}
    \centering
    \includegraphics[width=0.49\columnwidth]{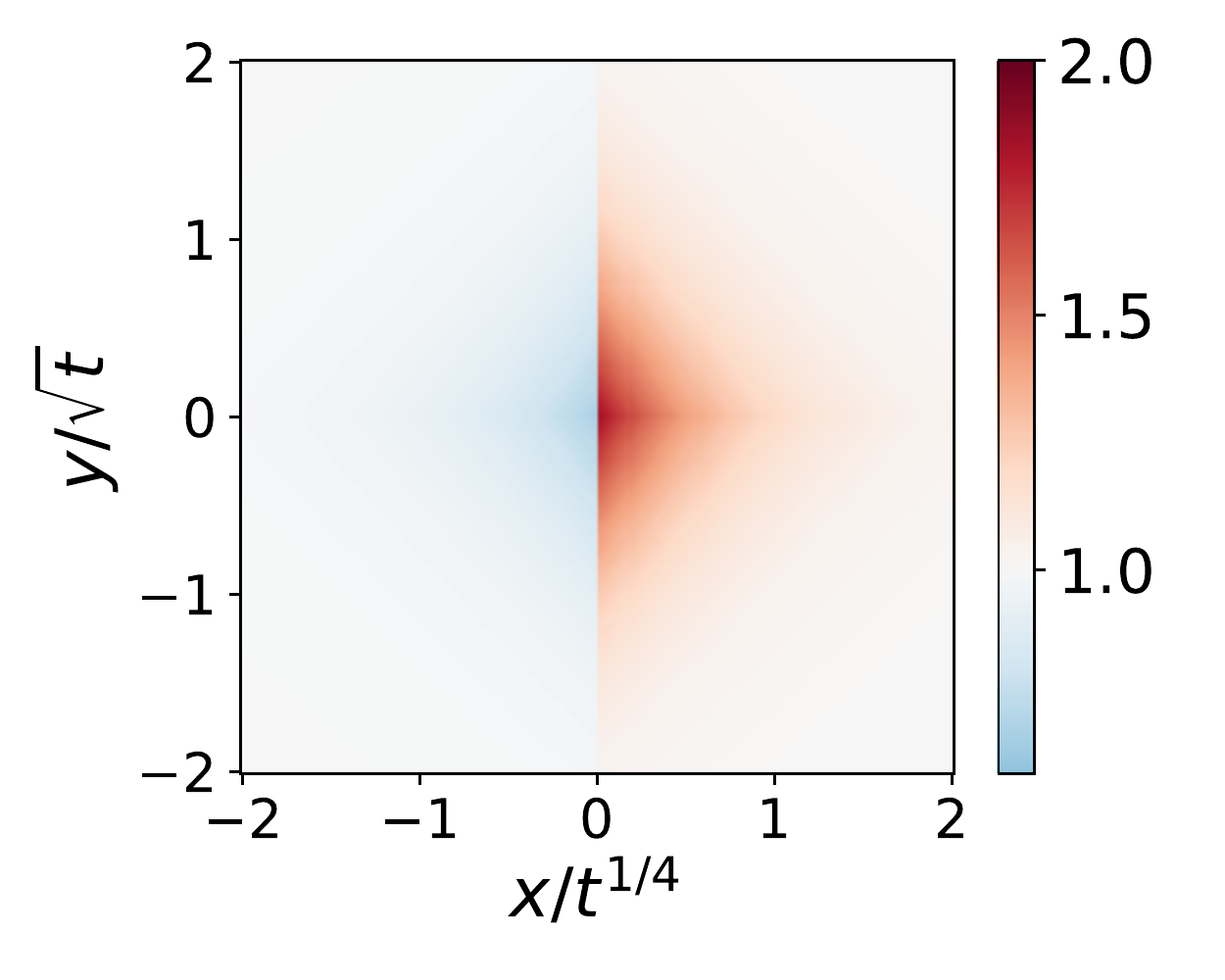}
    \includegraphics[width=0.49\columnwidth]{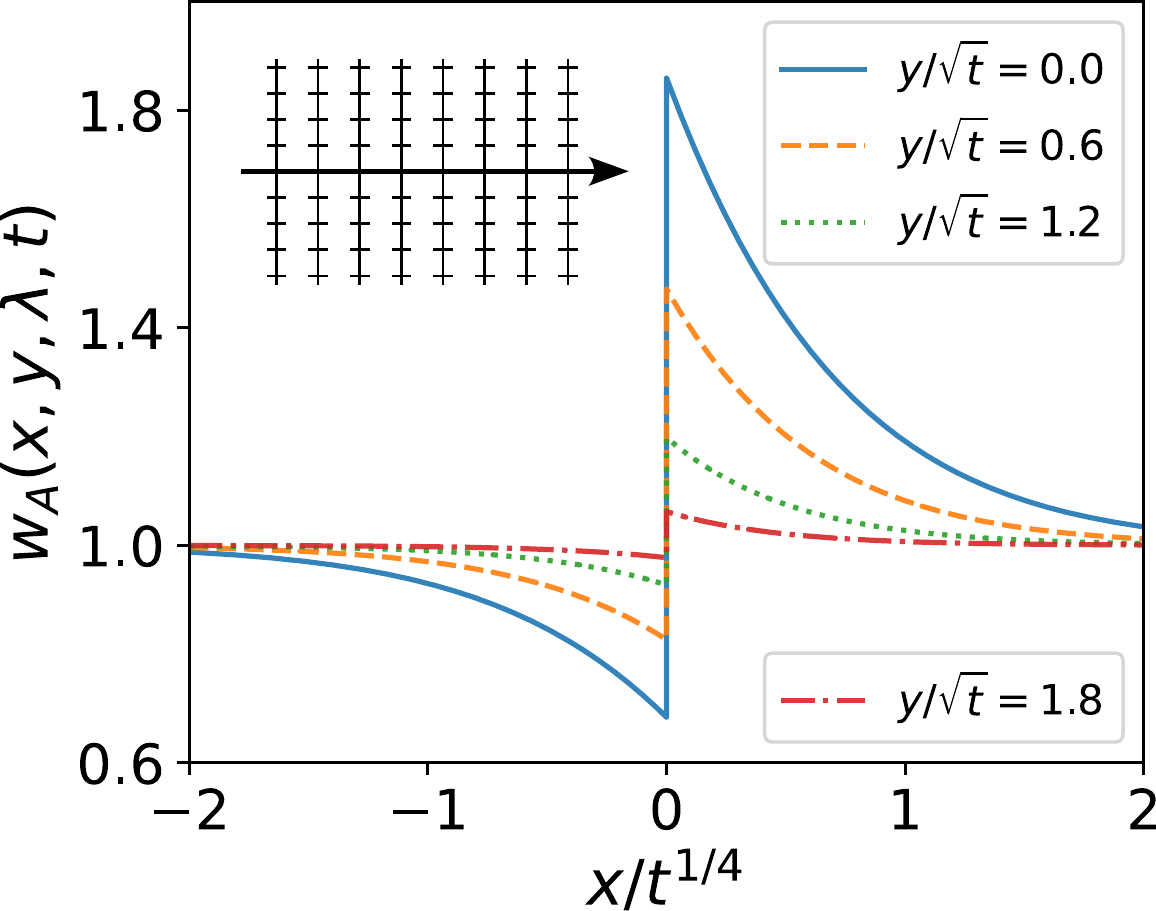}
    \caption{Correlation profile $\wA(\vec{r}=(x,y),\lambda,t)$ for random walkers on a comb lattice in the annealed case (represented on the right plot), in the continuous limit. Left: 2D representation. The profile is a scaling function of $x/t^{1/4}$ and $y/\sqrt{t}$. Right: Profile as a function of $x$ for fixed values of $y$.}
    \label{fig:ProfComb}
\end{figure}

\emph{Extension: other geometry.---} Our approach can be extended beyond the one dimensional case, in particular to any tree geometry. An important example which has generated lots of works is the comb lattice~\cite{Benichou:2015,Illien:2016,Iomin:2018}.
Comb structures have been developed to represent diffusion in critical percolation clusters, with the backbone and teeth of the comb mimicking the quasi-linear structure and the dead ends of percolation clusters~\cite{Weiss:1986}.
More recently, the comb model has been used to account for transport in real systems such as spiny dendrites~\cite{Mendez:2013}, diffusion of cold atoms~\cite{Sagi:2012} and diffusion in crowded media~\cite{Hofling:2013}.
It is a two dimensional lattice in which all the links parallel to the $x$-axis have been removed, except those on the axis itself, called the backbone (see the inset in Fig.~\ref{fig:ProfComb}). The propagator of a particle performing a random walk on this lattice is given in~\cite{Illien:2016}. In the continuous limit, the results~(\ref{eq:PsiAnnealedGenSF},\ref{eq:ProfAnnealedGenSF},\ref{eq:PsiQuenchGenSF},\ref{eq:ProfQuenchGenSF}) straightforwardly extend to this case, and leads to the correlation profile $\wA(\vec{r},\lambda,t)$ shown in Fig.~\ref{fig:ProfComb}. It presents a different scaling with time in the two directions $x$ and $y$, because particles can diffuse in the teeth of the comb, but horizontal motion is slowed down because it is only possible at $y=0$.

\emph{Extension: other observables.---}The above discussion can be extended to observables of the form
\begin{equation}
    \label{eq:GenObs}
  \mathcal{O}_t[f,g] = \sum_{i} \left[ f(x_i(t)) - g(x_i(0)) \right]
  \:,
\end{equation}
where $f$ and $g$ are two given functions, with $f(x) \underset{x \to \pm \infty}{\simeq} g(x)$ to ensure convergence of the sum. The case of the integrated current $Q_t$ corresponds to $f=g=\Theta$. In general, the above results~(\ref{eq:PsiAnnealedGenSF},\ref{eq:ProfAnnealedGenSF},\ref{eq:PsiQuenchGenSF},\ref{eq:ProfQuenchGenSF}) still hold, but with the tilted propagator
\begin{equation}
    \Gt{x}{y} = \e^{\lambda f(x)} G_t(x|y) \e^{-\lambda g(y)}
    \:.
\end{equation}
This provides in particular the profiles in the case of a generalised current $J_t(X)$, defined by $f(x) = \Theta(x-X)$ and $g(x) = \Theta(x)$. This observable measures the difference between the number of particles at the right of $X$ at time $t$ with the number of particles on the positive axis at $t=0$. It has proved to be especially relevant since it can be used to find the position $X_t$ of a tracer particle~\cite{Imamura:2017,Imamura:2021}, given by $J_t(X_t) = 0$, meaning that no particle can cross the tracer. However, this only provides the final position $X_t$, and not the displacement $X_t-X_0$ ($X_0$ being the position of the first particle to the right of the origin, which is random).

\emph{Extension: tracer particle.---} Nevertheless, our method can be adapted to directly study the displacement of a tracer (placed initially at the origin for simplicity).
So far, the only available studies concern the distribution of the tracer only~\cite{Rodenbeck:1998,Hegde:2014}, and not its correlations with the other particles, which are our main focus here.

\begin{figure}
    \centering
    \includegraphics[width=\columnwidth]{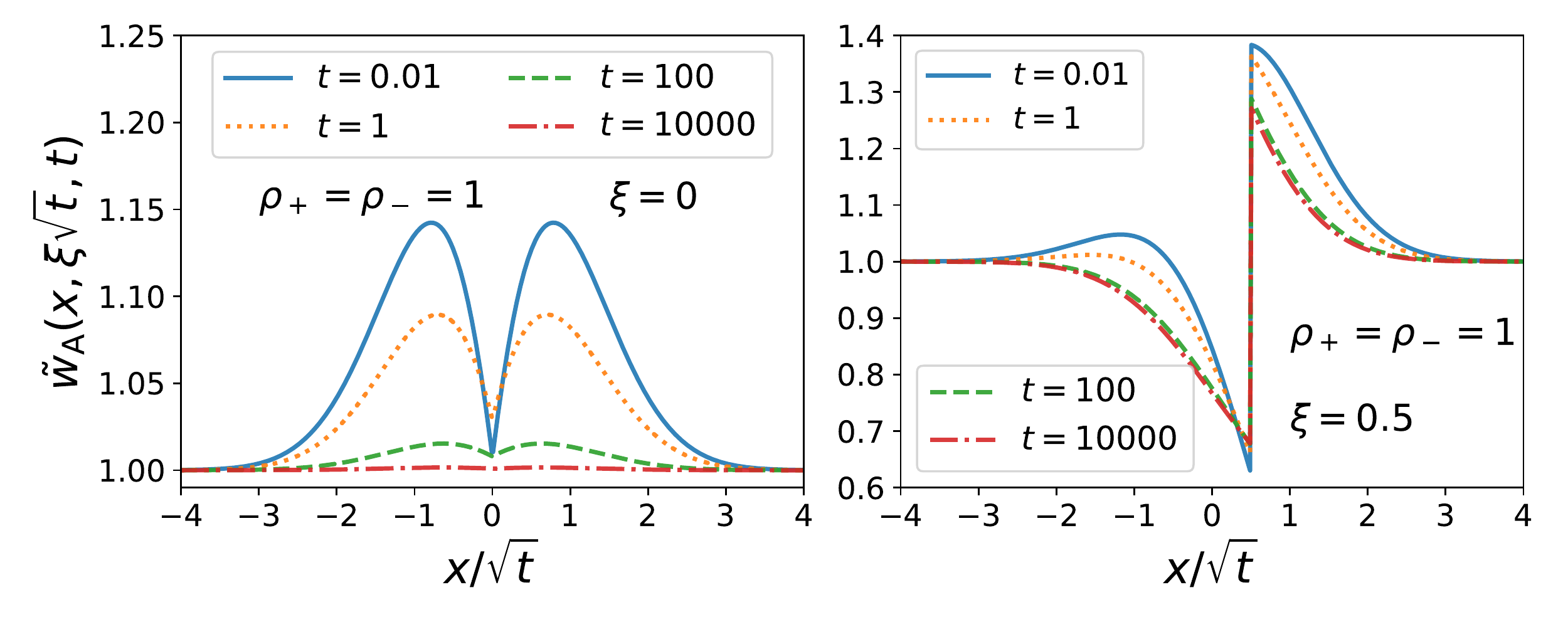}
    \caption{Conditional density profiles $\tilde{w}_\mathrm{A}$, for reflecting Brownian particles with diffusion coefficient $D=\frac{1}{2}$, with a tracer at position $X = \xi \sqrt{t}$, at different times $t$. Left: case $\xi=0$, corresponding to a tracer conditioned to be at its mean position. Right: $\xi=0.5$.}
    \label{fig:CondProfTr}
\end{figure}

Extending the ideas of~\cite{Levitt:1973,Rodenbeck:1998,Hegde:2014,Sadhu:2015,Krapivsky:2015a}, the correlation profiles can be computed by noticing that the tracer can be mapped onto the "middle" particle of the system of noninteracting particles introduced above.
We now consider a finite system of $2N+1$ particles, with initially $N$ particles on the negative axis (positions $x_{-n}(t)$), $N$ particles on the positive axis (positions $x_{n}(t)$), and a tracer initially at the origin ($x_0(t)$). We define the average over the initial positions as
\begin{multline}
    \label{eq:DefMoyItr}
    \moyI{f(\{ x_i(0)\})} = \int_{-\infty}^0 \prod_{n=1}^{N} \frac{\rho_0(y_{-n}) \dd y_{-n}}{N}
  \int_{0}^{\infty} \prod_{n=1}^{N} \frac{\rho_0(y_{n}) \dd y_{n}}{N} \\
  \: f(y_{-N}, \ldots, y_{-1}, y_0=0, y_1, \ldots y_N)
  \:.
\end{multline}
The average over the time evolution is still given by~\eqref{eq:DefMoyE}. The probability of finding the tracer at position $X$ at time $t$, with initially the particles at positions $\{ x_i(0) \}$ can be obtained by imposing that there are still $N$ particles to the left of the tracer, and $N$ to the right,
\begin{multline}
    \label{eq:PtX0}
    P_t(X|\{ x_i(0) \})
    \equiv \moyE{\delta(X-x_0(t))}
    \\
    \propto \int_{-\infty}^X\prod_{n=1}^N \dd x_{-n} \int_{X}^\infty \prod_{n=1}^N \dd x_n \:
    K_t( \left.\{ x_n \} \right|_{x_0=X} | \{ x_i(0) \})
    \:.
\end{multline}
Using the expression of the joint propagator~\eqref{eq:PropNpart}, and averaging over the time evolution and the initial positions (assumed to be annealed for simplicity), we obtain the distribution of $x_0(t)$ in the thermodynamic limit,
\begin{equation}
    \label{eq:PtX}
    P_t(X) \equiv \lim_{N \to \infty} \moyI{ \moyE{\delta(X-x_0(t))} }
    = \int_{-\pi}^\pi f_t(X,\theta) \e^{\phi_t(X,\theta)} \dd \theta
    \:,
\end{equation}
where the integration over $\theta$ enforces the noncrossing constraint. The functions $f_t$ and $\phi_t$  are expressed in terms of the propagator $G_t$ (assumed translationally invariant and symmetric), its integral $F_t(z) = \int_{z}^\infty G_t(x|0) \dd x$ and the initial density of particles $\rho_0$. The exact expressions are given in SM~\cite{SM}
The distribution~\eqref{eq:PtX} extends to the out-of-equilibrium case of an arbitrary initial density $\rho_0$, such as a step initial condition, the result of~\cite{Hegde:2014} obtained in the equilibrium case of a constant density. Here, we obtain in addition the full spatial dependence of the correlations between the position of the tracer and the density of surrounding particles, which takes the following simple and universal form
\begin{multline}
    \label{eq:ConfProfX}
    \tilde{w}_{\mathrm{A}}(x,X,t) \equiv \lim_{N \to \infty} \frac{\moyI{ \moyE{ \rho(x,t) \delta(X-x_0(t))} }}{  \moyI{ \moyE{\delta(X-x_0(t))} } }
    \\
    = \alpha^\pm_t(X) G_t(x|0) 
    + \beta^\pm_t(X) \rho_\mp F_t( \pm x) + \rho_\pm F_t(\mp x)
    \:,
\end{multline}
where the superscript $\pm$ stands for $x \gtrless X$, with $\alpha^\pm_t$ and $\beta^\pm_t$ given explicitly in SM~\cite{SM}. We stress that the spatial dependence of these profiles are fully encoded in the propagator $G_t$ (and its integral $F_t$).
Note that we have considered in Eq.~\eqref{eq:ConfProfX} here conditional profiles, which measure the mean density of particles conditioned on having observed the tracer at $X$ at time $t$. In contrast, we have previously considered correlation profiles $\moyI{\moyE{ \rho(x,t) \e^{\lambda x_0(t)} }}/\moyI{\moyE{\e^{\lambda x_0(t)} }}$. The two formulations are equivalent in the limit $t \to \infty$~\cite{Grabsch:2022,SM}, but not at arbitrary time. It turns out that the correlation profiles are more convenient to study observables of the form~\eqref{eq:GenObs}, while the conditional profiles are more suited to study tracer particles, since they take a simple form. These conditional profiles are shown in Fig.~\ref{fig:CondProfTr} for reflective Brownian particles. At long times, they reach their asymptotic values computed in~\cite{Poncet:2021}, but at arbitrary times they have a more complex structure, mostly due to the presence of the tracer at the origin at $t=0$ (first term in~\eqref{eq:ConfProfX}).

\emph{Extension: two tracers.---} Our approach can be extended to the important and largely unexplored situation of two tracers of positions $X_1(t)$ and $X_2(t)$. The only available studies in single-file systems concern the symmetric exclusion process and its limits~\cite{Majumdar:1991,Takeshi:2013,Ooshida:2018,Poncet:2018}, and the distance between two different particles at different times~\cite{Sabhapandit:2015} which does not gives access to the joint position of two tracers at the same time. We still obtain a simple form for the density profile conditioned on observing the first tracer at $X$ and the second tracer at $Y$. As a byproduct, we obtain the strikingly simple, universal, and to the best of our knowledge new expression for the covariance of the displacements of two tracers,
\begin{equation}
    \frac{\mathrm{Cov}(X_1,X_2)}{\sqrt{\mathrm{Var}(X_1) \mathrm{Var}(X_2)}}
    \underset{t \to \infty}{\simeq} 
    \frac{\int_z^\infty \dd x \int_{x}^\infty \dd y \: g(y)}
    {\int_0^\infty \dd x \int_{x}^\infty \dd y \: g(y)}
    \:,
\end{equation}
where $z = [X_1(0) - X_2(0)]/\sigma_t$, and $\sigma_t$ the long time scaling of the propagator $G_t(x|y) = g(\frac{x-y}{\sigma_t})/\sigma_t$.

\emph{Conclusion.---} We have shown that the correlation profiles in single-file systems take the strikingly simple universal form~(\ref{eq:ProfAnnealedGenSF}-\ref{eq:ProfQuenchGenSF}). 
The approach is general and applies to: (i) a broad range of dynamics (including with external forces or non-Markovian dynamics provided that the memory is lost upon collision); (ii) arbitrary time (note that these results are not accessible via the classical Macroscopic Fluctuation Theory~\cite{Bertini:2015}, which is limited to the large time behaviour of diffusive systems~\footnote{For diffusive systems, the MFT gives access to correlation profiles of the form $\moy{\rho(x,\tau T) \e^{\lambda \mathcal{O}_T}}/\moy{ \e^{\lambda \mathcal{O}_T} }$ with $T$ large and $0 \leq \tau \leq 1$. Here, we have obtained correlation profiles of the form $\moy{\rho(x,t) \e^{\lambda \mathcal{O}_t}}/\moy{ \e^{\lambda \mathcal{O}_t} }$ for arbitrary $t$.}); (iii)  different initial conditions (annealed and quenched~\footnote{The quenched case is notoriously more difficult than the annealed case: for instance while the annealed cumulant generating function of the current in the symmetric exclusion process has been determined in~\cite{Derrida:2009}, its quenched counterpart is still missing.}); (iv) various observables (the form~\eqref{eq:GenObs} includes the joint statistics of several currents~\cite{SM}); (v) geometries not restricted to the single-file constraint (illustrated by the comb geometry).
In addition, beyond the clear physical relevance of the correlation profiles, the simplicity of~(\ref{eq:ProfAnnealedGenSF}-\ref{eq:ProfQuenchGenSF}) further highlights their key role to describe transport properties in confined geometry~\cite{Krapivsky:2014,Imamura:2017,Imamura:2021,Poncet:2021,Grabsch:2022,Bettelheim:2022,Mallick:2022,Krajenbrink:2023,Grabsch:2023,Dean:2023}.


%

\let\addcontentsline\oldaddcontentsline

\clearpage
\widetext

\begin{center}
\begin{large}

    \textbf{
      Supplementary Material for\texorpdfstring{\\}{} From Particle Currents to Tracer Diffusion: Universal Correlation Profiles in Single-File Dynamics
    }
\end{large}
    
      \bigskip
      
      Aur{\'e}lien Grabsch, Th{\'e}otim Berlioz, Pierre Rizkallah, Pierre Illien, and Olivier B{\'e}nichou

  \end{center}

\setcounter{secnumdepth}{3}

\renewcommand{\theequation}{S\arabic{equation}}
\renewcommand{\thefigure}{S\arabic{figure}}

\setcounter{equation}{0}

\tableofcontents

\section{Annealed vs quenched correlation functions}

We consider the general observable
\begin{equation}
\label{eq:GenObsSM}
  \mathcal{O}_t[f,g] = \sum_{i} \left[ f(x_i(t)) - g(x_i(0)) \right]
  \:,
\end{equation}
with $f(x) \underset{x \to \pm \infty}{\simeq} g(x)$ to ensure convergence in the limit $N \to \infty$. We also define the density of particles
\begin{equation}
  \rho(x,t) = \sum_i \delta(x - x_i(t))
  \:.
\end{equation}
The joint statistical properties of these two observables are encoded in the joint cumulant generating functions. In the annealed case, it takes the form
\begin{equation}
    \log \moyI{ \moyE{ \e^{\lambda \Ot  + \chi \rho(x,t)}} }
    = \log \moyI{ \moyE{ \e^{\lambda \Ot } } } 
    + \chi \frac{\moyI{\moyE{\rho(x,t) \e^{\lambda \Ot}}}}{ \moyI{\moyE{\e^{\lambda \Ot}}}}
    + O(\chi^2)
    \:.
\end{equation}
Similarly, we consider the quenched joint cumulant generating function
\begin{equation}
    \moyI{ \log \moyE{ \e^{\lambda \Ot  + \chi \rho(x,t)}} }
    = \moyI{\log  \moyE{ \e^{\lambda \Ot } } } 
    + \chi \moyI{ \frac{\moyE{\rho(x,t) \e^{\lambda \Ot}}}{ \moyE{\e^{\lambda \Ot}}} }
    + O(\chi^2)
    \:.
\end{equation}
This motivates the definition of the correlation profiles $\wA$ and $\wQ$ in the main text.

\section{Derivation of the cumulants and correlations}
\label{sec:GenObs}

\subsection{Average over the time evolution}

We begin by computing the moment generating function for a given number $N$ of particles. We will denote $\{ y_i = x_i(0) \}$ the positions of the particles at initial time, which are at the moment treated as parameters. By definition,
\begin{align}
  \nonumber
  \moyE{\e^{\lambda \Ot}}
  &= \int_{-\infty}^\infty \dd x_1 \cdots \dd x_N \:
    K_t(\{ x_i \}| \{ y_i \}) \prod_{i=1}^N \e^{\lambda (f(x_i) - g(y_i))}
  \\
  \nonumber
  &= \frac{1}{N!} \sum_\sigma \prod_{i=1}^N \e^{- \lambda g(y_i)}
    \int_{-\infty}^\infty  G_t(x|y_{\sigma(i)}) \e^{\lambda f(x)} \dd x
  \\
  &= \prod_{i=1}^N \e^{-\lambda g(y_i)} \int_{-\infty}^\infty  G_t(x|y_i)
    \e^{\lambda f(x)} \dd x
  \:,
\end{align}
since all the integrals are independent and equivalent. Similarly, we can compute
\begin{align}
  \nonumber
  \moyE{\rho(x,t) \e^{\lambda \Ot}}
  &= \sum_{n} \int_{-\infty}^\infty \dd x_1 \cdots \dd x_N \: \delta(x-x_n)
    K_t(\{ x_i \}| \{ y_i \}) \prod_{i=1}^N \e^{\lambda (f(x_i) - g(y_i))}
  \\
  \nonumber
  &= \frac{1}{N!} \sum_n \sum_\sigma
    G_t(x|y_{\sigma(n)}) \e^{\lambda f(x) - \lambda g(y_n)}
    \prod_{i\neq n} \e^{- \lambda g(y_i)}
    \int_{-\infty}^\infty  G_t(x'|y_{\sigma(i)}) \e^{\lambda f(x')} \dd x'
  \\
  \nonumber
  &= \frac{1}{N!} \sum_n \sum_q \sum_{\sigma, \sigma(n) = q}
    G_t(x|y_{q}) \e^{\lambda f(x) - \lambda \sum_i g(y_i)}
    \prod_{j \neq q}
    \int_{-\infty}^\infty  G_t(x'|y_{j}) \e^{\lambda f(x')} \dd x'
  \\
  &= \sum_q
    G_t(x|y_{q}) \e^{\lambda f(x) - \lambda g(y_q)}
    \prod_{j \neq q}
    \int_{-\infty}^\infty  G_t(x'|y_{j}) \e^{\lambda f(x') - \lambda g(y_j)} \dd x'
  \:.
\end{align}
The higher order correlations with the density are also accessible in the same way, for instance,
\begin{align}
  \nonumber
  \moyE{\rho(x,t) \rho(y,t) \e^{\lambda \Ot}}
  &= \sum_{n,m} \int_{-\infty}^\infty \dd x_1 \cdots \dd x_N
    \: \delta(x-x_n) \delta(y-x_m)
    K_t(\{ x_i \}| \{ y_i \}) \prod_{i=1}^N \e^{\lambda (f(x_i) - g(y_i))}
  \\
  \nonumber
  & = \frac{1}{N!} \sum_{n \neq m} \sum_\sigma
    \tilde{G}_t(x|y_{\sigma(n)})
    \tilde{G}_t(y|y_{\sigma(m)})
    \prod_{i\neq {n,m}} \int_{-\infty}^\infty  \tilde{G}_t(x'|y_{\sigma(i)}) \dd x'
  \\
  \nonumber
  & + \frac{1}{N!} \sum_n \sum_\sigma
    \tilde{G}_t(x|y_{\sigma(n)}) \delta(x-y)
    \prod_{i\neq {n}} \int_{-\infty}^\infty  \tilde{G}_t(x'|y_{\sigma(i)}) \dd x'
  \\
  \nonumber
  &= \sum_{q \neq p} \tilde{G}_t(x|y_q) \tilde{G}_t(y|y_p)
    \prod_{j \neq {p,q}} \int_{-\infty}^\infty  \tilde{G}_t(x'|y_{j}) \dd x'
  \\
  & + \delta(x-y) \sum_q \tilde{G}_t(x|y_q)
    \prod_{j\neq {q}} \int_{-\infty}^\infty  \tilde{G}_t(x'|y_{j}) \dd x'
  \:,
\end{align}
where we have denoted $\tilde{G}_t(x|y) = \e^{\lambda f(x)} G_t(x|y) \e^{-\lambda g(y)}$. To take the average over the initial condition, we need to treat separately the annealed and quenched cases.

\subsection{Average over the initial positions in the annealed case}

In the annealed case, we compute
\begin{equation}
  \moyI{\moyE{\e^{\lambda \Ot}}}
  = \left( \frac{1}{N} \int \dd y \: \rho_0(y) \e^{-\lambda g(y)} \int G_t(x|y) \e^{\lambda f(x)} \dd x \right)^N
  \:.
\end{equation}
For finite $N$, the condition $\int \rho_0 = N$ ensures that the above
integrals are convergent. But in the limit $N \to \infty$ this would not be the case. Take for instance a constant density profile $\rho_0(x) = \rho$, then the integral would diverge in the domains $(x>0,y>0)$ and $(x<0,y<0)$. We can solve this issue by writing the above expression as
\begin{equation}
  \label{eq:MomGenFctAnnealedGenSF}
  \moyI{\moyE{\e^{\lambda \Ot}}}
  = \left(1+ \frac{1}{N} \int \dd y \: \rho_0(y) \int G_t(x|y) (\e^{\lambda f(x)- \lambda g(y)}-1) \dd x \right)^N
  \:,
\end{equation}
where we have used that $\int G_t(x|y)\dd x = 1$.
In this expression, if $x$ and $y$ are large, with the same sign, the
integrand vanishes. And if they are both large with opposite signs,
$G_t(x|y)$ is small. Therefore, the integral is always well defined,
and thus
\begin{equation}
  \label{eq:PsiAnnealedGenSFSM}
  \boxed{
  \psiA(\lambda,t) \equiv 
    \lim_{N \to \infty} \ln \moyI{\moyE{\e^{\lambda \Ot}}}
    =  \int \dd y \: \rho_0(y) \int G_t(x|y) (\e^{\lambda f(x)- \lambda g(y)}-1) \dd x
    \:.
  }
\end{equation}

To compute the density profiles, we need
\begin{equation}
  \moyI{\moyE{\rho(x,t) \e^{\lambda \Ot}}}
  =
  \left(
    \int \dd y \: \rho_0(y) G_t(x|y) \e^{\lambda f(x) - \lambda g(y)}
  \right)
  \left( \frac{1}{N} \int \dd y \: \rho_0(y) \e^{-\lambda g(y)} \int G_t(x'|y) \e^{\lambda f(x')} \dd x' \right)^{N-1}
  \:.
\end{equation}
Taking the ratio with~(\ref{eq:MomGenFctAnnealedGenSF}) and using the same rewriting as before, we get
\begin{equation}
  \frac{\moyI{\moyE{\rho(x,t) \e^{\lambda \Ot}}}}{ \moyI{\moyE{\e^{\lambda \Ot}}} }
  = \frac{ \e^{\lambda f(x)}  \int \dd y \: \rho_0(y) G_t(x|y) \e^{- \lambda g(y)}}
  {1 + \frac{1}{N} \int \dd y \: \rho_0(y) \int G_t(x'|y) (\e^{\lambda f(x')- \lambda g(y)}-1) \dd x'}
  \:.
\end{equation}
In the limit $N \to \infty$, this becomes
\begin{equation}
  \label{eq:ProfAnnealedGenSFSM}
  \boxed{
    \wA(x,\lambda,t) \equiv
    \lim_{N \to \infty}
    \frac{\moyI{\moyE{\rho(x,t) \e^{\lambda \Ot}}}}{ \moyI{\moyE{\e^{\lambda \Ot}}} }
    = \e^{\lambda f(x)}  \int \dd y \: \rho_0(y) G_t(x|y) \e^{- \lambda g(y)}
    \:.
  }
\end{equation}

For the two points correlation function, we need to compute
\begin{multline}
  \moyI{ \moyE{\rho(x,t) \rho(y,t) \e^{\lambda \Ot}} }
  =
  \\
  \sum_{q \neq p} \left(
    \int_{-\infty}^{\infty} \dd z \frac{\rho_0(z)}{N} \tilde{G}_t(x|z)
  \right) \left(
    \int_{-\infty}^{\infty} \dd z \frac{\rho_0(z)}{N} \tilde{G}_t(y|z)
  \right)
  \left(
    \int_{-\infty}^{\infty} \dd z \frac{\rho_0(z)}{N} \int_{-\infty}^\infty \dd x' \: \tilde{G}_t(x'|z)
  \right)^{N-2}
  \\
  + \delta(x-y) \sum_q \left(
    \int_{-\infty}^{\infty} \dd z \frac{\rho_0(z)}{N} \tilde{G}_t(x|z)
  \right)
  \left(
    \int_{-\infty}^{\infty} \dd z \frac{\rho_0(z)}{N} \int_{-\infty}^\infty \dd x' \: \tilde{G}_t(x'|z)
  \right)^{N-1}
  \:.
\end{multline}
The sums no longer depend on $p$ and $q$, and thus,
\begin{multline}
  \moyI{ \moyE{\rho(x,t) \rho(y,t) \e^{\lambda \Ot}} }
  =
  \\
  \frac{N-1}{N}
  \left(
    \int_{-\infty}^{\infty} \dd z \: \rho_0(z) \tilde{G}_t(x|z)
  \right) \left(
    \int_{-\infty}^{\infty} \dd z \: \rho_0(z) \tilde{G}_t(y|z)
  \right)
  \left(
    \int_{-\infty}^{\infty} \dd z \frac{\rho_0(z)}{N} \int_{-\infty}^\infty \dd x' \: \tilde{G}_t(x'|z)
  \right)^{N-2}
  \\
  + \delta(x-y) \left(
    \int_{-\infty}^{\infty} \dd z \: \rho_0(z) \tilde{G}_t(x|z)
  \right)
  \left(
    \int_{-\infty}^{\infty} \dd z \frac{\rho_0(z)}{N} \int_{-\infty}^\infty \dd x' \: \tilde{G}_t(x'|z)
  \right)^{N-1}
  \:.
\end{multline}
Taking the ratio with~(\ref{eq:MomGenFctAnnealedGenSF}), and using
again the same rewriting for the denominator, we get for $N \to \infty$,
\begin{multline}
  \lim_{N \to \infty}
  \frac{\moyI{ \moyE{\rho(x,t) \rho(y,t) \e^{\lambda \Ot}} }}
  { \moyI{ \moyE{ \e^{\lambda \Ot}} }  }
  =
  \delta(x-y) \left(
    \int_{-\infty}^{\infty} \dd z \: \rho_0(z) G_t(x|z)
    \e^{\lambda f(x) - \lambda g(z)}
  \right)
  \\
  + \left(
    \int_{-\infty}^{\infty} \dd z \: \rho_0(z) G_t(x|z)
    \e^{\lambda f(x) - \lambda g(z)}
  \right) \left(
    \int_{-\infty}^{\infty} \dd z \: \rho_0(z) G_t(y|z)
    \e^{\lambda f(x) - \lambda g(z)}
  \right)
  \:.
\end{multline}
In particular, we have
\begin{equation}
\boxed{
  \label{eq:TwoPointsCorrelAnnealGenSF}
  \lim_{N \to \infty} \left[ \frac{\moyI{ \moyE{\rho(x,t) \rho(y,t) \e^{\lambda \Ot}} }}
  { \moyI{ \moyE{ \e^{\lambda \Ot}} }  }
  -  \frac{\moyI{ \moyE{\rho(x,t) \e^{\lambda \Ot}} }}
  { \moyI{ \moyE{ \e^{\lambda \Ot}} }  }
  \frac{\moyI{ \moyE{\rho(y,t) \e^{\lambda \Ot}} }}
  { \moyI{ \moyE{ \e^{\lambda \Ot}} }  }
  \right]
  =
  \\
  \delta(x-y) \wA(x,\lambda,t)
  \:.
  }
\end{equation}

\subsection{Average over the initial positions in the quenched case}

In the quenched case, the cumulant generating function is given by
\begin{equation}
  \moyI{\ln \moyE{\e^{\lambda \Ot}}}
  =
  \sum_i \frac{1}{N} \int \dd y \:
  \rho_0(y) \ln \left[ \int_{-\infty}^\infty G_t(x|y) \e^{\lambda f(x)-\lambda g(y)}  \dd x \right]
  \:.
\end{equation}
The terms in the sum are independent of $i$, and the integral converges in the limit $N \to \infty$, thus
\begin{equation}
  \label{eq:PsiQuenchGenSFSM}
  \boxed{
  \psiQ(\lambda,t) \equiv
    \lim_{N \to \infty} \moyI{\ln \moyE{\e^{\lambda \Ot}}}
    = \int \dd y \:
    \rho_0(y) \ln \left[ \int_{-\infty}^\infty G_t(x|y) \e^{\lambda f(x)-\lambda g(y)}  \dd x \right]
    \:.
  }
\end{equation}
Similarly, for the profiles we compute
\begin{equation}
  \frac{ \moyE{\rho(x,t) \e^{\lambda \Ot}} }{  \moyE{ \e^{\lambda \Ot}} }
  = \sum_q
  \frac{G_t(x|y_{q}) \e^{\lambda f(x) - \lambda g(y_q)}}
  {\int_{-\infty}^\infty  G_t(x'|y_q) \e^{\lambda f(x') - \lambda g(y_q)} \dd x'}
  \:.
\end{equation}
We can easily take the average over the initial positions $y_i$, which gives
\begin{equation}
  \label{eq:ProfQuenchGenSFSM}
  \boxed{
  \wQ(x,\lambda,t) \equiv
    \lim_{N \to \infty}
    \moyI{ \frac{ \moyE{\rho(x,t) \e^{\lambda \Ot}} }{  \moyE{ \e^{\lambda \Ot}} } }
    = \int \dd y \: \rho_0(y)
    \frac{G_t(x|y) \e^{\lambda f(x) - \lambda g(y)}}
    {\int_{-\infty}^\infty  G_t(x'|y) \e^{\lambda f(x') - \lambda g(y)} \dd x'}
    \:.
  }
\end{equation}
Finally, for the two points correlation functions, we have
\begin{equation}
  \frac{ \moyE{\rho(x,t) \rho(y,t) \e^{\lambda \Ot}} }
  {  \moyE{ \e^{\lambda \Ot}} }
  = \sum_{q \neq p} \frac{\tilde{G}_t(x|y_p)}{\int_{-\infty}^{\infty} \tilde{G}_t(x'|y_p) \dd x'}
  \frac{\tilde{G}_t(y|y_q)}{\int_{-\infty}^{\infty} \tilde{G}_t(x'|y_q) \dd x'}
  \\
  + \delta(x-y) \sum_q  \frac{\tilde{G}_t(x|y_q)}{\int_{-\infty}^{\infty} \tilde{G}_t(x'|y_p) \dd x'}
  \:.
\end{equation}
Averaging over the initial condition yields
\begin{multline}
  \moyI{ \frac{ \moyE{\rho(x,t) \rho(y,t) \e^{\lambda \Ot}} } 
    {  \moyE{ \e^{\lambda \Ot}} } }
  =
  \delta(x-y) \int_{-\infty}^\infty \dd z \: \rho_0(z)
  \frac{\tilde{G}_t(x|z)}{\int_{-\infty}^{\infty} \tilde{G}_t(x'|z) \dd x'}
  \\
  + \frac{N-1}{N}
  \left( \int_{-\infty}^{\infty} \dd z \: \rho_0(z) 
    \frac{\tilde{G}_t(x|z)}{\int_{-\infty}^{\infty} \tilde{G}_t(x'|z) \dd x'}
  \right)
  \left(
    \int_{-\infty}^{\infty} \dd z \: \rho_0(z) 
    \frac{\tilde{G}_t(y|z)}{\int_{-\infty}^{\infty} \tilde{G}_t(x'|z) \dd x'}
  \right)
  \:.
\end{multline}
In the large $N$ limit, this gives
\begin{equation}
  \label{eq:TwoPointsCorrelQuenchGenSF}
  \boxed{
  \lim_{N \to \infty} \left[
  \moyI{ \frac{ \moyE{\rho(x,t) \rho(y,t) \e^{\lambda \Ot}}}
  {\moyE{ \e^{\lambda \Ot}} } }
  - \moyI{ \frac{ \moyE{\rho(x,t) \e^{\lambda \Ot}} }
    { \moyE{ \e^{\lambda \Ot}}  } }
  \: \moyI{
  \frac{\moyE{\rho(y,t) \e^{\lambda \Ot}}}
  { \moyE{ \e^{\lambda \Ot}}} }
  \right]
  =
   \delta(x-y) \wQ(x,\lambda,t)
  \:,
  }
\end{equation}
as in the annealed case.

\subsection{Applications}

In all the following applications, we consider an initial density of particles $\rho_0(x) = \rho_+ \Theta(x) + \rho_- \Theta(-x)$ for simplicity, and consider the case of the current $Q_t$, corresponding to $f(x) = g(x) = \Theta(x)$.

\subsubsection{Brownian particles}

As a first application, we revisit the well-studied case of reflecting Brownian particles on the real line. The propagator of an individual particle with diffusion coefficient $D$ is $G_t(x|y) = \e^{-\frac{(x-y)^2}{4Dt}}/\sqrt{4\pi Dt}$. Using this propagator in the general expressions~(\ref{eq:PsiAnnealedGenSFSM},\ref{eq:ProfAnnealedGenSFSM}), we obtain $\psiA(\lambda,t) = \omega \sqrt{\frac{D t}{\pi}}$ and
\begin{equation}
    \label{eq:wABrow}
    \wA(x>0,\lambda,t) =  \rho_+
    + \frac{\rho_- \e^{\lambda} - \rho_+}{2}  \erfc \left( \frac{x}{\sqrt{4Dt}} \right)
    \:,
\end{equation}
where we have denoted $\omega = \rho_+ (\e^{-\lambda}-1) + \rho_- (\e^{\lambda}-1)$, by analogy with the single parameter identified in the simple exclusion process (SEP)~\cite{Derrida:2009SM}. Indeed, these expressions coincide with the low density limit of the ones obtained for the SEP~\cite{Derrida:2009SM,Poncet:2021SM,Grabsch:2022SM}. The profile for $x<0$ can be deduced from the symmetry $x \to -x$, $\lambda \to -\lambda$, $\rho_\pm \to \rho_\mp$. Similarly, in the quenched case, we
recover the expression found in~\cite{Derrida:2009aSM} for the cumulants, and we additionally get the correlation profiles,
\begin{equation}
    \label{eq:wQBrow}
    \wQ(x,\lambda,t) = 
     \e^{\lambda \Theta(x)}
    \int_{-\infty}^\infty \rho_0(y) \frac{\e^{-\left(\frac{x}{\sqrt{4Dt}}-y\right)^2}}{1 + \frac{\e^\lambda-1}{2} \erfc(-y)} 
    \frac{\dd y}{\sqrt{\pi}}
    \:.
\end{equation}
A related expression for constant $\rho_0$ can be found in~\cite{Krapivsky:2015SM}. The two profiles $\wA$ and $\wQ$ are represented in Fig.~\ref{fig:ProfLevy} (solid black lines).

\subsubsection{Run and tumble particles}

The Laplace transform of the propagator of an individual particle takes a simple form~\cite{Weiss:2002SM},
\begin{equation}
    \int_0^\infty \e^{-s t} G_t(x|y) \dd t 
    = \frac{\Lambda(s)}{2s} \e^{-\Lambda(s) \abs{x-y}}
    \:,
\end{equation}
where $\Lambda(s) = \sqrt{s(s+2\gamma)}/v_0$. We can easily obtain the annealed profile and cumulant generating function in the Laplace domain since the expressions~(\ref{eq:PsiAnnealedGenSFSM},\ref{eq:ProfAnnealedGenSFSM}) are linear in the propagator. The inverse Laplace transform can be computed explicitly using the expressions given in~\cite{Banerjee:2020SM}, and we get
$\psiA(\lambda,t) = \frac{\omega}{2} v_0 t \: \e^{-\gamma t}
(\mathrm{I}_0(\gamma t) + \mathrm{I}_1(\gamma t))$, where we have again denoted $\omega = \rho_+ (\e^{-\lambda}-1) + \rho_- (\e^{\lambda}-1)$, by analogy with the single parameter identified in the simple exclusion process (SEP)~\cite{Derrida:2009SM}, and $\mathrm{I}_\nu$ is a modified Bessel function. Similarly, the correlation profile reads
\begin{equation}
    \label{eq:wARTPSM}
    \wA(x > 0,\lambda,t) =
    \rho_+
  + \frac{\rho_- \e^\lambda - \rho_+}{2}
  \Theta(v_0 t - x)
  \left(
    \e^{-\frac{\gamma x}{v_0}}
    + \frac{\gamma x}{v_0}
    \int_1^{\frac{v_0 t}{x}} \frac{\e^{-\frac{\gamma x T}{v_0}}
      \mathrm{I}_1(\frac{\gamma x}{v_0} \sqrt{T^2-1})}{\sqrt{T^2-1}}
    \dd T
  \right)
  \:.
\end{equation}
The expression for $x<0$ can again be obtained by symmetry. Finally, in the large time limit, $\psiA$ and $\wA$~\eqref{eq:wARTPSM} reduce to the Brownian case, with a diffusion coefficient $D=v_0^2/(2\gamma)$.

\subsubsection{Lévy flights}

\begin{figure}
    \centering
    \includegraphics[width=0.8\columnwidth]{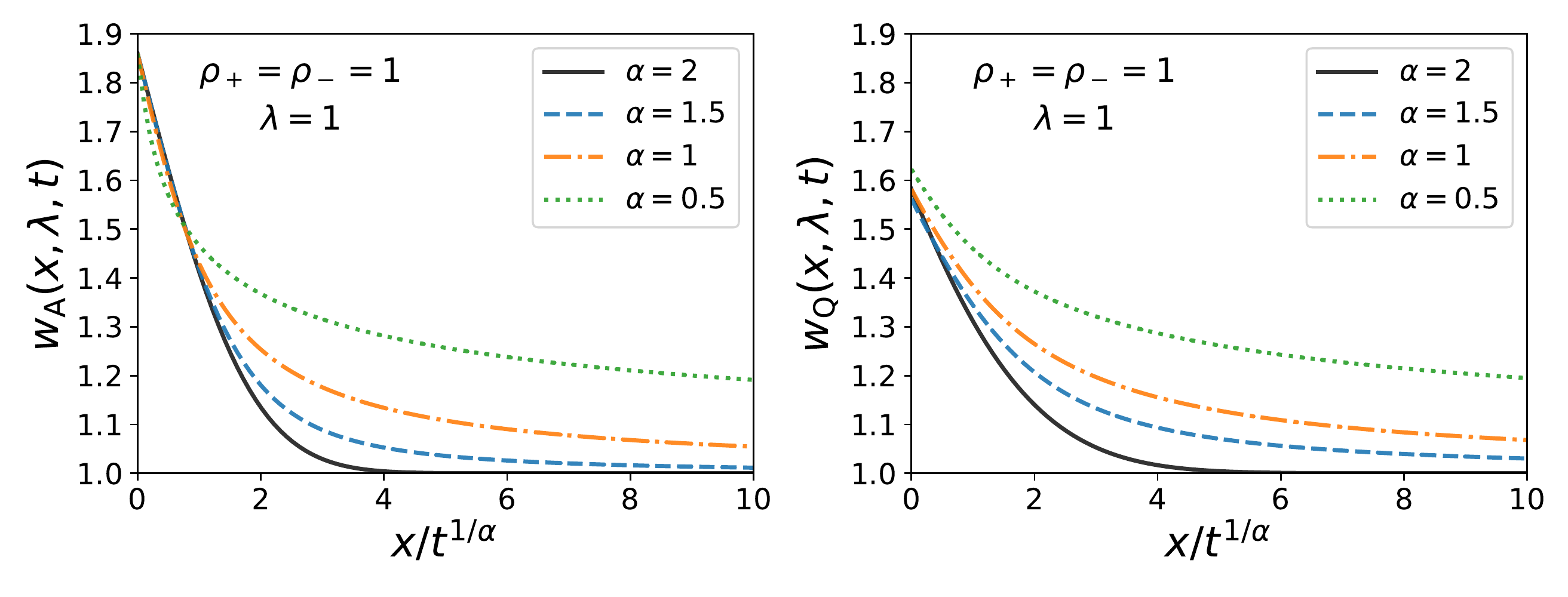}
    \caption{Annealed correlation profile $\wA$ (left) and quenched correlation profile $\wQ$ (right) for Lévy particles, for different values of the Lévy exponent $\alpha$. They are compared to the one of the Brownian particles with $D=1$, corresponding to $\alpha = 2$ (black solid line). For $0 < \alpha < 2$, the correlation profiles decay as a power law, as $x^{-\alpha}$, while for the Brownian particles, the decay is much faster, as $\e^{-\frac{x^2}{4Dt}}/x$.}
    \label{fig:ProfLevy}
\end{figure}

Next, as an illustration of non diffusive dynamics at long times, we consider particles performing Lévy flights, corresponding to an individual propagator
\begin{equation}
    G_t(x|y) = \frac{1}{t^{1/\alpha}} g \left( \frac{x-y}{t^{1/\alpha}} \right)
    \:,
    \quad
    g(x) = \int_{-\infty}^\infty \e^{-\I k x - \abs{k}^\alpha} \frac{\dd k}{2 \pi}
    \:,
\end{equation}
with the exponent $\alpha \in ]0,2]$.
Plugging this expression into the formulas~\eqref{eq:ProfAnnealedGenSFSM} for $\wA$ and~\eqref{eq:ProfQuenchGenSFSM} for $\wQ$, we obtain that these two profiles are functions of $x/t^{1/\alpha}$ only. The annealed profile reads
\begin{equation}
    \wA(x = z t^{1/\alpha},\lambda,t) = \rho_+ 
    + (\rho_- \e^\lambda - \rho_+) \int_{z}^\infty g(y) \dd y
    \:,
\end{equation}
for $x > 0$. Similarly, the quenched profile is
\begin{equation}
    \wQ(x = z t^{1/\alpha},\lambda,t) = \e^{\lambda \Theta(z)} \left[ \rho_+
    \int_0^\infty \dd y \frac{g(z-y)}{1 + (\e^{\lambda}-1) \int_{-y}^\infty g(u) \dd u}
    + \rho_- \int_{-\infty}^0 \dd y \frac{g(z-y)}{1 + (\e^{\lambda}-1) \int_{-y}^\infty g(u) \dd u}
    \right]
    \:.
\end{equation}
They are represented for different values of $\alpha$ in Fig.~\ref{fig:ProfLevy}. The correlations present a power law behaviour, $\wA(x,\lambda,t)-\rho_+ \underset{x \to \infty}{\sim} x^{-\alpha}$ for $0<\alpha < 2$, due to the fact that the particles can perform large jumps in this case.

\section{The comb geometry}

\subsection{Propagator in the continuous limit}

The comb lattice is a two dimensional lattice in which all the links parallel to the $x$-axis have been removed, except those on the axis itself, called the backbone. The propagator $P_t(\vec{r}|\vec{s})$ of a particle performing a random walk in discrete time on this lattice is given in~\cite{Illien:2016SM}. Its expressions depends is different if the arrival point $\vec{r} = (r_x,r_y)$ and the starting point $\vec{s} = (r_x',r_y')$ are on the same teeth of the comb ($r_x=r_x'$ and $r_y r_y' >0$) or not. The (discrete) Laplace transform of the propagator reads~\cite{Illien:2016SM}
\begin{equation}
    \label{eq:PropCombDiffTooth}
    \hat{P}_\xi(\vec{r}|\vec{s}) 
    \equiv \sum_{t=0}^\infty \xi^t P_t(\vec{r}|\vec{s})
    = \frac{1 + \delta_{r_y,0}}{2} G_2(\xi) f_2(\xi)^{\abs{r_x-r_x'}} f_1(\xi)^{\abs{r_y}+\abs{r_y'}}
    \:,
    \quad
    \text{if}
    \quad
    r_x \neq r_x' \text{ or } r_y r_y' < 0
    \:,
\end{equation}
\begin{equation}
    \label{eq:PropCombSameTooth}
    \hat{P}_\xi(\vec{r}|\vec{s}) =
    f_1(\xi)^{\abs{r_y-r_y'}} \left\lbrace
    \frac{1}{2} G_2(\xi) f_1(\xi)^{2 \check{r}}
    + \frac{2}{\xi} f_1(\xi)^{2\check{r}-1}
    + G_1(\xi) \left[
    1 - f_1(\xi)^{2(\check{r} - 1)}
    \right]
    \right\rbrace
    \:,
    \quad
    \text{if}
    \quad
    r_x = r_x' \text{ and } r_y r_y' > 0
    \:,
\end{equation}
where $\check{r} = \min(\abs{r_y}, \abs{r_y'})$, and
\begin{equation}
    f_1(\xi) = \frac{1 - \sqrt{1-\xi^2}}{\xi}
    \:,
    \quad
    f_2(\xi) = \frac{1 + \sqrt{1-\xi^2} - \sqrt{2}\sqrt{1-\xi^2 + \sqrt{1-\xi^2} } }{\xi}
    \:,
\end{equation}
\begin{equation}
    G_1(\xi) = \frac{1}{\sqrt{1-\xi^2}}
    \:,
    \quad
    G_2(\xi) = \sqrt{\frac{2}{1-\xi^2 + \sqrt{1-\xi^2}}}
    \:.
\end{equation}

One can define a continuous version of this random walk in different ways. A convenient one is to define it as the long time limit of the discrete one. We introduce a scaling parameter $T$ and appropriately rescale space and time with $T$ and let $T \to \infty$. We obtain a nontrivial limit with the following scalings,
\begin{equation}
    r_x = x \: T^{1/4}
    \:,
    \quad
    r_y = y \sqrt{T}
    \:,
    \quad
    t = \tau \: T
    \:,
\end{equation}
with $x$, $y$ and $\tau$ fixed. This implies $t \to \infty$ when $T \to \infty$, corresponding to $\xi \to 1$, such that $1-\xi = u/T$ in the Laplace domain. Inserting these scaling into the propagator~\eqref{eq:PropCombDiffTooth}, we get,
\begin{equation}
    \hat{P}_\xi(\vec{r}|\vec{r'})
    \underset{T \to \infty}{\simeq} \frac{1}{2} \left( \frac{2 T}{u} \right)^{1/4} 
    (1 - 2^{3/4} u^{1/4} T^{-1/4} )^{\abs{x-x'} T^{1/4}}
    (1 - \sqrt{2 u}/\sqrt{T})^{(\abs{y} + \abs{y'}) \sqrt{T}}
    \:.
\end{equation}
We can define the propagator in the continuous limit as
\begin{equation}
    \tilde{P}_u(x,y|x',y') \equiv \lim_{T \to \infty} T^{-1/4} \hat{P}_\xi(\vec{r}|\vec{r'})
    = \frac{1}{2} \left( \frac{2}{u} \right)^{1/4} 
    \e^{-2^{3/4} u^{1/4} \abs{x-x'}- \sqrt{2u}(\abs{y} + \abs{y'})}
    \:,
\end{equation}
where the prefactor $T^{-1/4} = T^{1/4} \times \sqrt{T} \times ( T )^{-1}$ comes from the rescaling of both space directions and time (and thus $\xi$). This expression holds for different teeth of the comb, thus $x \neq x'$. Taking the continuous limit of~\eqref{eq:PropCombSameTooth} for the case of $x=x'$ (same teeth), we get
\begin{equation}
    \tilde{P}_u(x,y|x' = x,y') \equiv \lim_{T \to \infty} T^{-1/2} \hat{P}_\xi(\vec{r}|\vec{r'})
    = \frac{1}{\sqrt{2u}} \e^{- \sqrt{2u} \abs{y-y'}}
    ( 1 - \e^{-2 \sqrt{2u} \min(\abs{y},\abs{y'})} )
    \:,
\end{equation}
where this time the prefactor $T^{-1/2} = \sqrt{T} \times \left( T \right)^{-1}$ comes from the rescaling of $y$ and $t$ (and not $x$ since it is fixed). We can combine these two expression into
\begin{equation}
    \label{eq:CombPropContLaplace}
    \tilde{P}_u(x,y|x',y') 
    = 
    \frac{1}{2^{3/4}u^{1/4}}
    \e^{-2^{3/4} u^{1/4} \abs{x-x'}- \sqrt{2u}(\abs{y} + \abs{y'})}
    +
    \frac{1}{\sqrt{2u}} \left( \e^{- \sqrt{2u} \abs{y-y'}}
     - \e^{- \sqrt{2u} \abs{y+y'})} \right)\delta(x-x') \Theta(yy')
    \:,
\end{equation}
where $\Theta$ is the Heaviside step function.
Note that the second term is the Laplace transform of a propagator of a random walk on $y$ with reflecting boundary condition at the origin. One can check that this propagator is indeed normalised by integrating over $x$ and $y$,
\begin{equation}
    \int_{-\infty}^\infty \dd x \int_{-\infty}^\infty \dd y \: \tilde{P}_u(x,y|x',y') 
    = \frac{1}{u}
    \:,
\end{equation}
which is indeed the Laplace transform of $1$.

\subsection{Current through the bond \texorpdfstring{$(0,0)-(1,0)$}{(0,0)-(1,0)}}

As in the 1D case, the current through the horizontal bond $(0,0)-(1,0)$ can be expressed in terms of the number of particles on the right half space (this is true because the lattice is a tree). In the continuous limit, $Q_t$ is thus again given by~\eqref{eq:GenObsSM}, with $f(x) = g(x) = \Theta(x)$ as in the main text. The derivation of Section~\ref{sec:GenObs} can be straightforwardly adapted to the case of the comb by replacing the 1D integrals by two-dimensional integrals. For instance, the annealed correlation profile reads
\begin{equation}
    \wA(x,y,\lambda,\tau) = \e^{\lambda \Theta(x)} \int \dd x' \int \dd y' \rho_0(x',y') P_\tau(x,y|x',y') \e^{-\lambda \Theta(x')}
    \:.
\end{equation}
Since this expression is linear in the propagator, one can take its Laplace transform in time,
\begin{equation}
    \tilde{w}_{\mathrm{A}}(x,y,\lambda,u) \equiv 
    \int_{0}^\infty \wA(x,y,\lambda,\tau) \e^{-u \tau} \dd \tau
    = \e^{\lambda \Theta(x)} \int \dd x' \int \dd y' \rho_0(x',y') \tilde{P}_u(x,y|x',y') \e^{-\lambda \Theta(x')}
    \:.
\end{equation}
Computing these integrals with the propagator~\eqref{eq:CombPropContLaplace}, with a constant density $\rho_0(x,y) = \rho$, we obtain,
\begin{equation}
    \tilde{w}_{\mathrm{A}}(x,y,\lambda,u) = \frac{\rho}{u}
    \left(1
    +  \frac{\e^{\lambda \: \sg{x}}-1}{2} \e^{-2^{3/4} u^{1/4} \abs{x} - \sqrt{2 u} \abs{y} }
    \right)
    \:.
\end{equation}
This expression can be inverted to the time domain numerically, and in plotted in the main text. The fact that $\tilde{w}_{\mathrm{A}}$ is a function of $u^{1/4} x$ and $\sqrt{u} y$ in the Laplace domain implies that $\wA$ is a function of $x/\tau^{1/4}$ and $y/\sqrt{\tau}$ in the time domain.

\section{Derivation of the cumulants and correlations for the tracer}

\subsection{Averaging over the time evolution}

We consider that we have initially $2N+1$ particles, at positions $y_n$, $n \in [-N,N]$. We take the middle particle to be the
tracer, initially at position $y_0=0$. Therefore, $y_{-n} < 0$ for $n>0$ and $y_n >0$ for $n>0$. The distribution $P_t(X|\{ y_i \})$ of the tracer at time $t$ is obtained by imposing that $N$ particles remain on the left of the tracer, and $N$ on the right,
\begin{equation}
  P_t(X|\{ y_i \})
  \equiv \moyE{ \delta(X-x_0(t)) }
  = (2N+1) \binom{2N}{N} \prod_{n=1}^N \int_{-\infty}^X \dd x_{-n} \int_{X}^\infty \dd x_n \: K_t(\{ x_i \}|\{ y_i \})
  \:,
\end{equation}
where $K_t$ is now the propagator of the $2N+1$ particles, and we set $x_0 = X$. The combinatorial factor comes from the fact that the tracer can be any of the (2N+1) particles, and then we choose $N$ particles among the remaining $2N$ to be on the left of $X$. This gives,
\begin{align}
  P_t(x|\{ y_i \})
  &= \frac{1}{(N!)^2} \sum_{\sigma}
  G_t(X|y_{\sigma(0)})
  \prod_{n=1}^N \left( \int_{-\infty}^X \dd z \: G_t(z|y_{\sigma(-n)}) \right)
    \left(\int_{X}^{\infty} \dd z \: G_t(z|y_{\sigma(n)}) \right)
  \\
  &= \frac{1}{(N!)^2} \sum_{\sigma}
  G_t(X|y_{\sigma(0)})
  \prod_{n=1}^N \left( \int_{-X}^\infty \dd z \: G_t(-z|y_{\sigma(-n)}) \right)
    \left(\int_{X}^{\infty} \dd z \: G_t(z|y_{\sigma(n)}) \right)
  \:.
\end{align}
Partitioning the summations over the value of $\sigma(0)$, we get,
\begin{equation}
  \label{eq:DistrXt00}
  P_t(X|\{ y_i \}) =
  \frac{1}{(N!)^2} \sum_{q=-N}^N \sum_{\sigma, \sigma(0) = q}
  G_t(X|y_{q})
  \prod_{n=1}^N \left( \int_{0}^\infty \dd z \: G_t(-z|y_{\sigma(-n)}-X) \right)
    \left(\int_{0}^{\infty} \dd z \: G_t(z|y_{\sigma(n)}-X) \right)
  \:.
\end{equation}
In the product, all the terms are present, except $y_q$. The signs of $z$ and $n$ are distributed so that there are $N$ positive and $N$ negative ones (representing the particles on the right and on the left of the tracer respectively). Introducing $\varepsilon_n = \pm 1$ to represent these sign variables, we obtain
\begin{equation}
  \label{eq:DistrXt0}
  P_t(X|\{ y_i \}) = \sum_{n=-N}^N G_t(X|y_n)
  \sum_{\varepsilon_j = \pm 1} \delta_{\sum_{j \neq n} \varepsilon_j,0}
  \prod_{i \neq n} \int_{0}^\infty G_t(\varepsilon_i z | y_i-X) \dd z
  \:.
\end{equation}
with the Kroenecker $\delta$ enforcing that there are $N$ minus signs, and $N$ plus signs. The $(N!)^2$ term cancelled with the fact that~\eqref{eq:DistrXt00} is invariant under the permutation of the $N$ particles on the left and the $N$ particles on the right. Representing the Kroenecker $\delta$ as
\begin{equation}
  \delta_{x,y} = \frac{1}{2\pi} \int_{-\pi}^\pi \dd \theta \: \e^{i (x-y) \theta}
  \:,
\end{equation}
we get
\begin{equation}
  \label{eq:DistrXtGenProp0}
    P_t(X|\{ y_i \}) = \int_{-\pi}^\pi \frac{\dd \theta}{2\pi}
    \sum_{n=-N}^N G_t(X|y_n)
    \prod_{j \neq n} \int_{0}^\infty \left(
      \e^{\I \theta} G_t(z|y_i-X) + \e^{-\I \theta} G_t(-z|y_i-X)
    \right)
    \dd z
    \:.
\end{equation}

We can proceed similarly to study the average density of particles
with a tracer at position $X$:
\begin{equation}
  \moyE{\rho(x,t) \delta(x_0-X)} = \sum_{q \neq 0} \moyE{\delta(x-x_q) \delta(X-x_0)}
  = 
  \frac{(2N+1)!}{(N!)^2}
  \sum_{q \neq 0}  \prod_{n=1}^N \int_{-\infty}^X \dd x_{-n} \int_{X}^\infty \dd x_n \: K_t(\{ x_i \}|\{ y_i \}) \delta(x-x_q)
  \:.
\end{equation}
Splitting the sum over $q>0$ and $q<0$, we get
\begin{multline}
  \moy{\rho(x,t) \delta(x_0-X)}_E =
  \\
  \frac{ \Theta(X-x)}{(N!)^2}\sum_{q<0} \sum_\sigma G_t(X|y_{\sigma(0)}) G_t(x|y_{\sigma(q)})
  \prod_{n=-N, n\neq q}^{-1} \left( \int_{-\infty}^X \dd z \: G_t(z|y_{\sigma(n)}) \right)
  \prod_{n=1}^N
  \left(\int_{X}^{\infty} \dd z \: G_t(z|y_{\sigma(n)}) \right)
  \\+
  \frac{ \Theta(x-X)}{(N!)^2}\sum_{q>0} \sum_\sigma G_t(X|y_{\sigma(0)}) G_t(x|y_{\sigma(q)})
  \prod_{n=-N}^{-1} \left( \int_{-\infty}^X \dd z \: G_t(z|y_{\sigma(n)}) \right)
  \prod_{n=1,n\neq q}^N
  \left(\int_{X}^{\infty} \dd z \: G_t(z|y_{\sigma(n)}) \right)
  \:.
\end{multline}
As we did for the tracer only, we can get rid of the sum over the permutations by labelling the permuted indices $\sigma(n) = j$, and introducing auxilliary variables $\varepsilon_j = \pm 1$ to keep track of the signs which originate from the integrals to the left of $X$ or to the right of $X$. This gives
\begin{multline}
  \moyE{\rho(x,t) \delta(x_0-X)} =
  \\
  \Theta(X-x)
  \sum_{n=-N}^N \sum_{p\neq n} G_t(X|y_n) G_t(x|y_p)
  \sum_{\{\varepsilon_j\}, \varepsilon_p = -1} \delta_{\sum_{j \neq n} \varepsilon_j,0}
  \prod_{i \neq n,p} \int_{0}^\infty G_t(\varepsilon_i z | y_i-X) \dd z
  \\+
  \Theta(x-X)
  \sum_{n=-N}^N \sum_{p\neq n} G_t(X|y_n) G_t(x|y_p)
  \sum_{\{\varepsilon_j\}, \varepsilon_p = +1} \delta_{\sum_{j \neq n} \varepsilon_j,0}
  \prod_{i \neq n,p} \int_{0}^\infty G_t(\varepsilon_i z | y_i-X) \dd z
  \:.
\end{multline}
Introducing again the integral representation of the Kroenecker delta, we obtain
\begin{multline}
  \label{eq:JointDensXtArbPropag0}
  \moyE{\rho(x,t) \delta(x_0-X)} =
  \int_{-\pi}^\pi \frac{\dd \theta}{2\pi}\left(
    \e^{-\I \theta} \Theta(X-x) + \e^{\I \theta} \Theta(x-X)
  \right)
  \sum_{n=-N}^N G_t(X|y_n) \sum_{p \neq n} G_t(x|y_p)
  \\
  \times
  \prod_{j \neq n,p} \int_{0}^\infty \left(
    \e^{\I \theta} G_t(z|y_i-X) + \e^{-\I \theta} G_t(-z|y_i-X)
  \right)
  \dd z
  \:.
\end{multline}
We can easily check that, by integrating over $x$, we recover $2N$ times~(\ref{eq:DistrXtGenProp0}), as it should.

\subsection{Average over the initial positions}

We now average over the initial positions
of the particles. We consider that the particles are distributed
according to a density $\rho_0$, such that
\begin{equation}
  \int_{0}^\infty \rho_0(y) \dd y = N
  \:,
  \quad
  \int_{-\infty}^0 \rho_0(y) \dd y = N
  \:.
\end{equation}
On each side of the tracer, the particles are indistinguishable, so we
write the average over the positions $y_n$ as
\begin{equation}
  \moyI{(\cdots)} = \int_{-\infty}^0 \prod_{n=1}^{N} \frac{\rho_0(y_{-n}) \dd y_{-n}}{N}
  \int_{0}^{\infty} \prod_{n=1}^{N} \frac{\rho_0(y_{n}) \dd y_{n}}{N} \: (\cdots)
  \:.
\end{equation}
We can thus compute
\begin{multline}
  \moyI{P_t(X|\{ y_i \})} = \int_{-\pi}^\pi \frac{\dd \theta}{2\pi}
  \left\lbrace
    G_t(X,0) 
    \left[\int_{-\infty}^0 \frac{\dd y}{N} \rho_0(y) \tilde{G}_t(X,y,\theta)  \right]^{N}
    \left[\int_{0}^\infty \frac{\dd y}{N} \rho_0(y) \tilde{G}_t(X,y,\theta)  \right]^{N}
  \right.
  \\
  \sum_{n=-N}^{-1} \int_{-\infty}^0 \frac{\dd y_{-n}}{N} \rho_0(y_{-n}) G_t(X|y_{-n})
  \left[\int_{-\infty}^0 \frac{\dd y}{N} \rho_0(y) \tilde{G}_t(X,y,\theta)  \right]^{N-1}
  \tilde{G}_t(X,0,\theta)
  \left[\int_{0}^\infty \frac{\dd y}{N} \rho_0(y) \tilde{G}_t(X,y,\theta)  \right]^{N}
  \\
  \left.
    +
    \sum_{n=1}^{N} \int_{0}^\infty \frac{\dd y_{n}}{N} \rho_0(y_{n}) G_t(X|y_{n})
    \left[\int_{-\infty}^0 \frac{\dd y}{N} \rho_0(y) \tilde{G}_t(X,y,\theta)  \right]^{N}
    \left[\int_{0}^\infty \frac{\dd y}{N} \rho_0(y) \tilde{G}_t(X,y,\theta)  \right]^{N-1}
    \tilde{G}_t(X,0,\theta)
  \right\rbrace
  \:,
\end{multline}
with
\begin{equation}
  \tilde{G}_t(X,y,\theta) = \int_{0}^\infty \left(
    \e^{\I \theta} G_t(z|y-X) + \e^{-\I \theta} G_t(-z|y-X)
  \right)
  \dd z
  \:.
\end{equation}
We can write this expression in a more compact form, as
\begin{multline}
  \moyI{P_t(X|\{ y_i \})} = \int_{-\pi}^\pi \frac{\dd \theta}{2\pi}
  \left[\int_{-\infty}^0 \frac{\dd y}{N} \rho_0(y) \tilde{G}_t(X,y,\theta)  \right]^{N}
    \left[\int_{0}^\infty \frac{\dd y}{N} \rho_0(y) \tilde{G}_t(X,y,\theta)  \right]^{N}
  \Bigg\lbrace
  G_t(X,0)
  \\
  +
  N  \tilde{G}_t(X,0,\theta) \frac{\int_{-\infty}^0 \dd y \: \rho_0(y) G_t(X|y)}
  {\int_{-\infty}^0 \dd y \: \rho_0(y) \tilde{G}_t(X,y,\theta)}
  + N  \tilde{G}_t(X,0,\theta) \frac{\int_{0}^\infty \dd y \: \rho_0(y) G_t(X|y)}
  {\int_{0}^\infty \dd y \: \rho_0(y) \tilde{G}_t(X,y,\theta)}
  \Bigg\rbrace
  \:.
\end{multline}
In order to take the limit $N \to \infty$, we need to ensure the
convergence of the integrals. We thus rewrite
\begin{equation}
  \int_{-\infty}^0 \frac{\dd y}{N} \rho_0(y) \tilde{G}_t(X,y,\theta)
  =
  \e^{-\I \theta} + \frac{\e^{\I \theta} - \e^{-\I \theta}}{N} \int_{-\infty}^0 \dd y \: \rho_0(y) \int_{0}^\infty
  \dd z \: G_t(z|y-X)
  \:,
\end{equation}
\begin{equation}
  \int_{0}^\infty \frac{\dd y}{N} \rho_0(y) \tilde{G}_t(X,y,\theta)
  =
  \e^{\I \theta} + \frac{\e^{-\I \theta} - \e^{\I \theta}}{N} \int_{0}^\infty \dd y \: \rho_0(y) \int_{0}^\infty
  \dd z \: G_t(-z|y-X)
  \:,
\end{equation}
which are convergent for $N \to \infty$. We thus get,
\begin{multline}
  \lim_{N \to \infty} \moy{P_t(X|\{ y_i \})}_I
  =
  \int_{-\pi}^\pi \frac{\dd \theta}{2\pi}
  \Bigg\lbrace
  G_t(X,0)
  \\
  + \tilde{G}_t(X,0,\theta)
  \left[ \e^{\I \theta}\int_{-\infty}^0 \dd y \: \rho_0(y) G_t(X|y)
    +  \e^{-\I \theta} \int_{0}^\infty \dd y \: \rho_0(y) G_t(X|y)
  \right]
  \Bigg\rbrace
  \\
   \exp\left[
    (\e^{2\I \theta}-1) \int_{-\infty}^0 \dd y \: \rho_0(y) \int_{0}^\infty
    \dd z \: G_t(z|y-X)
  \right.
  \left.
    + (\e^{-2\I \theta}-1 )\int_{0}^\infty \dd y \: \rho_0(y) \int_{0}^\infty
    \dd z \: G_t(-z|y-X)
  \right]
  \:.
\end{multline}
Let us write this expression in a more compact form, as
\begin{equation}
\boxed{
  \label{eq:DistrXtallTimes}
  P_t(X) = \lim_{N \to \infty}
  \moyI{P_t(X|\{ y_i \})}
  = \int_{-\pi}^{\pi} \frac{\dd \theta}{2 \pi} f_{X,t}(\theta) \: \e^{\phi_{X,t}(\theta)}
  \:,
 }
\end{equation}
with
\begin{align}
  \phi_{X,t}
  &=
    (\e^{2\I \theta}-1) \int_{-\infty}^0 \dd y \: \rho_0(y) \int_{0}^\infty
    \dd z \: G_t(z|y-X)
    + (\e^{-2\I \theta}-1 )\int_{0}^\infty \dd y \: \rho_0(y) \int_{0}^\infty
    \dd z \: G_t(-z|y-X)
    \:,
  \\
  f_{X,t}(\theta)
  &=
    G_t(X,0)
    + \tilde{G}_t(X,0,\theta)
    \left[ \e^{\I \theta}\int_{-\infty}^0 \dd y \: \rho_0(y) G_t(X|y)
    +  \e^{-\I \theta} \int_{0}^\infty \dd y \: \rho_0(y) G_t(X|y)
    \right]
    \:.
\end{align}

This coincides with the expression given in~\cite{Hegde:2014SM} for constant initial density $\rho_0$. Proceeding similarly from~(\ref{eq:JointDensXtArbPropag0}), we get
\begin{multline}
  \moyI{\moyE{\rho(x,t) \delta(x_0-X)}} =
  \int_{-\pi}^\pi \frac{\dd \theta}{2\pi}\left(
    \e^{-\I \theta} \Theta(X-x) + \e^{\I \theta} \Theta(x-X)
  \right)
  \\
  \times
  \left[\int_{-\infty}^0 \frac{\dd y}{N} \rho_0(y) \tilde{G}_t(X,y,\theta)  \right]^{N}
  \left[\int_{0}^\infty \frac{\dd y}{N} \rho_0(y) \tilde{G}_t(X,y,\theta)  \right]^{N}
  \\
  \Bigg\lbrace
  G_t(X,0)
  \left[
    N \frac{\int_{-\infty}^0 \dd y \: \rho_0(y) G_t(x|y)}
    {\int_{-\infty}^0 \dd y \: \rho_0(y) \tilde{G}_t(X,y,\theta)}
    + N \frac{\int_{0}^\infty \dd y \: \rho_0(y) G_t(x|y)}
    {\int_{0}^\infty \dd y \: \rho_0(y) \tilde{G}_t(X,y,\theta)}
  \right]
  \\
  + G_t(x,0) \left[
    N \frac{\int_{-\infty}^0 \dd y \: \rho_0(y) G_t(X|y)}
    {\int_{-\infty}^0 \dd y \: \rho_0(y) \tilde{G}_t(X,y,\theta)}
    + N \frac{\int_{0}^\infty \dd y \: \rho_0(y) G_t(X|y)}
    {\int_{0}^\infty \dd y \: \rho_0(y) \tilde{G}_t(X,y,\theta)}
  \right]
  \\
 +N \tilde{G}_t(X,0,\theta) \frac{\int_{-\infty}^0 \dd y \: \rho_0(y) G_t(X|y)}
  {\int_{-\infty}^0 \dd y \: \rho_0(y) \tilde{G}_t(X,y,\theta)}
  \left[
    (N-1) \frac{\int_{-\infty}^0 \dd y \: \rho_0(y) G_t(x|y)}
    {\int_{-\infty}^0 \dd y \: \rho_0(y) \tilde{G}_t(X,y,\theta)}
    + N \frac{\int_{0}^\infty \dd y \: \rho_0(y) G_t(x|y)}
    {\int_{0}^\infty \dd y \: \rho_0(y) \tilde{G}_t(X,y,\theta)}
  \right]
  \\
  + N \tilde{G}_t(X,0,\theta) \frac{\int_{0}^\infty \dd y \: \rho_0(y) G_t(X|y)}
  {\int_{0}^\infty \dd y \: \rho_0(y) \tilde{G}_t(X,y,\theta)}
  \left[
    N \frac{\int_{-\infty}^0 \dd y \: \rho_0(y) G_t(x|y)}
    {\int_{-\infty}^0 \dd y \: \rho_0(y) \tilde{G}_t(X,y,\theta)}
    + (N-1) \frac{\int_{0}^\infty \dd y \: \rho_0(y) G_t(x|y)}
    {\int_{0}^\infty \dd y \: \rho_0(y) \tilde{G}_t(X,y,\theta)}
  \right]
  \Bigg\rbrace
  \:.
\end{multline}
Using the same large $N$ analysis as before, we get
\begin{multline}
  \label{eq:jointProfAllTime}
  \lim_{N \to \infty} \moyI{\moyE{\rho(x,t) \delta(x_0-X)}}
  =
  \int_{-\pi}^\pi \frac{\dd \theta}{2\pi} \left(
    \e^{-\I \theta} \Theta(X-x) + \e^{\I \theta} \Theta(x-X)
  \right)
  \\
   \exp\left[
    (\e^{2\I \theta}-1) \int_{-\infty}^0 \dd y \: \rho_0(y) \int_{0}^\infty
    \dd z \: G_t(z|y-X)
  \right.
  \left.
    + (\e^{-2\I \theta}-1 )\int_{0}^\infty \dd y \: \rho_0(y) \int_{0}^\infty
    \dd z \: G_t(-z|y-X)
  \right]
  \\
  \Bigg\lbrace
  \left[ G_t(X,0)
    + \tilde{G}_t(X,0,\theta) \left(
          \e^{\I \theta} \int_{-\infty}^0 \dd y \: \rho_0(y) G_t(X|y)
    + \e^{-\I \theta} \int_{0}^\infty \dd y \: \rho_0(y) G_t(X|y)
    \right)
  \right]
  \\
  \times
  \left[
    \e^{\I \theta} \int_{-\infty}^0 \dd y \: \rho_0(y) G_t(x|y)
    + \e^{-\I \theta} \int_{0}^\infty \dd y \: \rho_0(y) G_t(x|y)
  \right]
  \\
  + G_t(x,0) \left[
    \e^{\I \theta} \int_{-\infty}^0 \dd y \: \rho_0(y) G_t(X|y)
    + \e^{-\I \theta} \int_{0}^\infty \dd y \: \rho_0(y) G_t(X|y)
  \right]
  \Bigg\rbrace
  \:.
\end{multline}
Isolating the dependence in $x$, we can write the conditional profile in a compact form as
\begin{equation}
    \boxed{
  \lim_{N \to \infty} \frac{ \moyI{\moyE{\rho(x,t) \delta(x_0-X)}}}{ \moyI{\moyE{ \delta(x_0-X)}} }
  = \alpha_t^\pm(X) \: G_t(x|0)
  + \beta_t^\pm(X) \int_{\mathbb{R}^\mp} \rho_0(y) G_t(x|y) \dd y
  + \int_{\mathbb{R}^\pm} \rho_0(y) G_t(x|y) \dd y
  \:,
  }
\end{equation}
with the $\pm$ signs corresponding to $x \gtrless X$
\begin{equation}
  \alpha^\pm_t(X)
  = \frac{
    \int_{-\pi}^\pi \dd \theta \: \e^{\pm \I \theta} g_{X,t}(\theta) \: \e^{\phi_{X,t}(\theta)}
  }{
    \int_{-\pi}^\pi \dd \theta \: f_{X,t}(\theta) \: \e^{\phi_{X,t}(\theta)}
  }
  \:,
  \quad
  \beta_t^\pm(X)
  = \frac{
    \int_{-\pi}^\pi \dd \theta \: \e^{\pm 2 \I \theta} f_{X,t}(\theta) \: \e^{\phi_{X,t}(\theta)}
    }{
    \int_{-\pi}^\pi \dd \theta \: f_{X,t}(\theta) \: \e^{\phi_{X,t}(\theta)}
    }
    \:,
\end{equation}
with
\begin{equation}
  g_{X,t}(\theta) =
   \e^{\I \theta} \int_{-\infty}^0 \dd y \: \rho_0(y) G_t(X|y)
   + \e^{-\I \theta} \int_{0}^\infty \dd y \: \rho_0(y) G_t(X|y)
   \:.
\end{equation}

For a step initial density profile, $G_t$ invariant by translation and symmetric, these expressions reduce to the ones given in the main text.

\section{The case of two tracers}

Let us consider the joint distribution of the positions of two tracers, denoted $x_0(t)$ and $x_K(t)$, with $x_0(t) < x_K(t)$ without loss of generality. We assume that there are initially $K-1$ particles between the two tracers, at positions $x_n(t)$, $1 \leq n < K$. Initially, we have $x_0(0) = 0$, and $x_K(0) = Y_0$. As in the previous case, we place $N$ particles on the left of $x_0$, denoted $x_{-n}$ for $0 < n \leq N$, and $N$ particles on the right of $x_K$, denoted $x_n$ with $K < n \leq N+K$. We consider that these particles are initially distributed according to a density $\rho_0$ such that
\begin{equation}
    \int_{-\infty}^0 \rho_0 = N
    \:,
    \quad
    \int_{0}^{Y_0} \rho_0 = K-1
    \:,
    \quad
    \int_{Y_0}^\infty \rho_0 = N
    \:.
\end{equation}
We will compute the averages over the time evolution of the particles, and their initial positions for finite $N$, and then let $N \to \infty$.

\subsection{Averaging over the time evolution}

The computations are similar to the case of one tracer, except that we must now enforce two constraints: the number $N$ of particles on the left of $x_0$ is fixed, and the number $N$ of particles to the right of $x_K$ is also fixed. The total number of particles being conserved by the joint propagator, this automatically enforces that there are $K-1$ particles between $x_0$ and $x_K$. We again denote $\{ y_i \}$ the initial positions of the particles. We get,
\begin{align}
    \label{eq:PtTwoX}
    P_t(X,Y|\{ y_i \}) 
    &\equiv \moyE{\delta(X-x_0(t)) \delta(Y - x_K(t))}
    \\
    &= \int_{-\pi}^\pi \frac{\dd \theta}{2\pi} \e^{-\I N \theta}
    \int_{-\pi}^\pi \frac{\dd \varphi}{2\pi} \e^{-\I N \varphi}
    \sum_{n=-N}^{N+K} G_t(X|y_n) \sum_{m \neq n} G_t(Y|y_m)
    \prod_{j \neq n,m} \left(I_{\mathrm{L}}(y_j) \e^{\I \theta}
    + I_{\mathrm{C}}(y_j)
    + I_{\mathrm{R}}(y_j) \e^{\I \varphi}
    \right)
    \:,
    \nonumber
\end{align}
where we have defined
\begin{equation}
    I_{\mathrm{L}}(y) = \int_{-\infty}^X G_t(z|y) \dd z
    \:,
    \quad
    I_{\mathrm{C}}(y) = \int_{X}^Y G_t(z|y) \dd z
    \:,
    \quad
    I_{\mathrm{R}}(y) = \int_{Y}^\infty G_t(z|y) \dd z
    \:.
\end{equation}
The integrals over the two phases in~\eqref{eq:PtTwoX} enforce the two conservation constraints. Eq.~\eqref{eq:PtTwoX} is the analogous to Eq.~\eqref{eq:DistrXtGenProp0} obtained for one tracer. This reasoning can be extended to an arbitrary number of tracers. The same procedure can be used to obtain
\begin{multline}
    \label{eq:JointTrDens}
    \moyE{\delta(X-x_0(t)) \delta(Y - x_K(t)) \rho(x,t)}
    = 
    \int_{-\pi}^\pi \frac{\dd \theta}{2\pi} \e^{-\I N \theta}
    \int_{-\pi}^\pi \frac{\dd \varphi}{2\pi} \e^{-\I N \varphi}
    (\e^{\I \theta} \Theta(X-x) + \Theta(x-X)\Theta(Y-x) + \e^{\I \varphi} \Theta(x-Y) )
    \\
    \sum_{p} G_t(x|y_p) \sum_{n \neq p} G_t(X|y_n) \sum_{m \neq n,p} G_t(Y|y_m)
    \prod_{j \neq p,n,m} \left(I_{\mathrm{L}}(y_j) \e^{\I \theta}
    + I_{\mathrm{C}}(y_j)
    + I_{\mathrm{R}}(y_j) \e^{\I \varphi}
    \right)
    \:.
\end{multline}

\subsection{Averaging over the initial condition}

We must now compute the average over the initial positions $\{ y_i \}$, defined as
\begin{equation}
    \moyI{f(\{ y_i \})} = \int_{-\infty}^0 \prod_{n=1}^N \frac{\dd y_{-n}}{N}
     \int_{0}^{Y_0} \prod_{n=1}^K \frac{\dd y_{n}}{K-1}
     \int_{Y_0}^{\infty} \prod_{n=K+1}^{K+N} \frac{\dd y_{n}}{N}
     f(y_{-N}, \ldots, y_{-1},0, y_1, \ldots, y_{K-1}, Y_0, y_{K+1},\ldots,y_{N+K})
     \:.
\end{equation}
Performing this averaging for the probability~\eqref{eq:PtTwoX}, we obtain,
\begin{multline}
    P_t(X,Y) \equiv
    \moyI{P_t(X,Y|\{ y_i \})} =
    \int_{-\pi}^\pi \frac{\dd \theta}{2\pi} \e^{-\I N \theta}
    \int_{-\pi}^\pi \frac{\dd \varphi}{2\pi} \e^{-\I N \varphi}
    \left[ J_{\mathrm{L}}  \right]^N 
    \left[ J_{\mathrm{C}}  \right]^{K-1}
    \left[ J_{\mathrm{R}}  \right]^N
    \Bigg\lbrace
    G_t(X|0) G_t(Y|Y_0)
    + G_t(X|Y_0) G_t(Y|0)
    \\
    + G_t(X|0) \left[
        \frac{\int_{-\infty}^0 \rho_0(y) G_t(Y|y) \dd y}{J_{\mathrm{L}}}
        + \frac{\int_{0}^{Y_0} \rho_0(y) G_t(Y|y) \dd y}{J_{\mathrm{C}}}
        + \frac{\int_{Y_0}^\infty \rho_0(y) G_t(Y|y) \dd y}{J_{\mathrm{R}}}
    \right] \left[I_{\mathrm{L}}(Y_0) \e^{\I \theta}
    + I_{\mathrm{C}}(Y_0)
    + I_{\mathrm{R}}(Y_0) \e^{\I \varphi}
    \right]
    \\
    + G_t(X|Y_0) \left[
        \frac{\int_{-\infty}^0 \rho_0(y) G_t(Y|y) \dd y}{J_{\mathrm{L}}}
        + \frac{\int_{0}^{Y_0} \rho_0(y) G_t(Y|y) \dd y}{J_{\mathrm{C}}}
        + \frac{\int_{Y_0}^\infty \rho_0(y) G_t(Y|y) \dd y}{J_{\mathrm{R}}}
    \right]\left[I_{\mathrm{L}}(0) \e^{\I \theta}
    + I_{\mathrm{C}}(0)
    + I_{\mathrm{R}}(0) \e^{\I \varphi}
    \right]
    \\
    + G_t(Y|0) \left[
        \frac{\int_{-\infty}^0 \rho_0(y) G_t(Y|y) \dd y}{J_{\mathrm{L}}}
        + \frac{\int_{0}^{Y_0} \rho_0(y) G_t(Y|y) \dd y}{J_{\mathrm{C}}}
        + \frac{\int_{Y_0}^\infty \rho_0(y) G_t(Y|y) \dd y}{J_{\mathrm{R}}}
    \right] \left[I_{\mathrm{L}}(Y_0) \e^{\I \theta}
    + I_{\mathrm{C}}(Y_0)
    + I_{\mathrm{R}}(Y_0) \e^{\I \varphi}
    \right]
    \\
    + G_t(Y|Y_0) \left[
        \frac{\int_{-\infty}^0 \rho_0(y) G_t(Y|y) \dd y}{J_{\mathrm{L}}}
        + \frac{\int_{0}^{Y_0} \rho_0(y) G_t(Y|y) \dd y}{J_{\mathrm{C}}}
        + \frac{\int_{Y_0}^\infty \rho_0(y) G_t(Y|y) \dd y}{J_{\mathrm{R}}}
    \right]\left[I_{\mathrm{L}}(0) \e^{\I \theta}
    + I_{\mathrm{C}}(0)
    + I_{\mathrm{R}}(0) \e^{\I \varphi}
    \right]
    \\
    + \left[I_{\mathrm{L}}(0) \e^{\I \theta}
    + I_{\mathrm{C}}(0)
    + I_{\mathrm{R}}(0) \e^{\I \varphi}
    \right]
    \left[I_{\mathrm{L}}(Y_0) \e^{\I \theta}
    + I_{\mathrm{C}}(Y_0)
    + I_{\mathrm{R}}(Y_0) \e^{\I \varphi}
    \right] \times \Bigg[
    \\
    \frac{\int_{-\infty}^0 \rho_0(y)G_t(X|y) \dd y}{J_{\mathrm{L}}}
    \left(
        \frac{N-1}{N} \frac{\int_{-\infty}^0 \rho_0(y)G_t(Y|y) \dd y}{J_{\mathrm{L}}}
        + \frac{\int_{0}^{Y_0} \rho_0(y) G_t(Y|y) \dd y}{J_{\mathrm{C}}}
        + \frac{\int_{Y_0}^\infty \rho_0(y) G_t(Y|y) \dd y}{J_{\mathrm{R}}}
    \right)
    \\
    + \frac{\int_{0}^{Y_0}\rho_0(y)G_t(Y|y) \dd y}{J_{\mathrm{C}}}
        \left(
        \frac{\int_{-\infty}^0 \rho_0(y)G_t(Y|y) \dd y}{J_{\mathrm{L}}}
        + \frac{K-2}{K-1}\frac{\int_{0}^{Y_0} \rho_0(y)G_t(Y|y) \dd y}{J_{\mathrm{C}}}
        + \frac{\int_{Y_0}^\infty \rho_0(y)G_t(Y|y) \dd y}{J_{\mathrm{R}}}
    \right)
    \\
    +  \frac{\int_{Y_0}^\infty \rho_0(y)G_t(Y|y) \dd y}{J_{\mathrm{R}}}
    \left(
        \frac{\int_{-\infty}^0 \rho_0(y)G_t(Y|y) \dd y}{J_{\mathrm{L}}}
        + \frac{\int_{0}^{Y_0} \rho_0(y) G_t(Y|y) \dd y}{J_{\mathrm{C}}}
        + \frac{N-1}{N} \frac{\int_{Y_0}^\infty \rho_0(y) G_t(Y|y) \dd y}{J_{\mathrm{R}}}
    \right)
    \Bigg]
    \Bigg\rbrace
\end{multline}
where
\begin{equation}
    J_{\mathrm{L}} = \int_{-\infty}^0 \frac{\dd y}{N} \rho_0(y) \left(I_{\mathrm{L}}(y) \e^{\I \theta}
    + I_{\mathrm{C}}(y)
    + I_{\mathrm{R}}(y) \e^{\I \varphi}
    \right)
    \:,
    \quad
    J_{\mathrm{R}} = \int_{Y_0}^\infty \frac{\dd y}{N} \rho_0(y) \left(I_{\mathrm{L}}(y) \e^{\I \theta}
    + I_{\mathrm{C}}(y)
    + I_{\mathrm{R}}(y) \e^{\I \varphi}
    \right)
    \:,
\end{equation}
\begin{equation}
    J_{\mathrm{C}} = \int_{0}^{Y_0} \frac{\dd y}{K-1} \rho_0(y) \left(I_{\mathrm{L}}(y) \e^{\I \theta}
    + I_{\mathrm{C}}(y)
    + I_{\mathrm{R}}(y) \e^{\I \varphi}
    \right)
    \:.
\end{equation}
To take the thermodynamic limit $N \to \infty$, we must regularise the expressions of $J_{\mathrm{L}}$ and $J_{\mathrm{R}}$ as
\begin{equation}
    J_{\mathrm{L}} = \e^{\I \theta} + \int_{-\infty}^0 \frac{\dd y}{N} \rho_0(y) \left[(I_{\mathrm{L}}(y)-1) \e^{\I \theta}
    + I_{\mathrm{C}}(y)
    + I_{\mathrm{R}}(y) \e^{\I \varphi}
    \right]
    \:,
\end{equation}
\begin{equation}
    J_{\mathrm{R}} = \e^{\I \varphi} + \int_{Y_0}^\infty \frac{\dd y}{N} \rho_0(y) \left[I_{\mathrm{L}}(y) \e^{\I \theta}
    + I_{\mathrm{C}}(y)
    + (I_{\mathrm{R}}(y)-1) \e^{\I \varphi}
    \right]
    \:.
\end{equation}
Taking the limit $N \to \infty$ yields
\begin{equation}
    \label{eq:PtXY}
    \boxed{
    \lim_{N \to \infty} P_t(X,Y) = \int_{-\pi}^\pi \frac{\dd \theta}{2\pi} 
    \int_{-\pi}^\pi \frac{\dd \varphi}{2\pi}
    f_t(X,Y,Y_0,\theta,\varphi) \e^{\phi_t(X,Y,Y_0,\theta,\varphi)}
    \:,
    }
\end{equation}
where
\begin{multline}
    \phi_t(X,Y,\theta,\varphi)
    = \int_{-\infty}^0 \dd y \: \rho_0(y) \left[(I_{\mathrm{L}}(y)-1)
    + I_{\mathrm{C}}(y) \e^{-\I \theta}
    + I_{\mathrm{R}}(y) \e^{\I (\varphi-\theta)}
    \right]
    \\
    + \int_{Y_0}^\infty \dd y \: \rho_0(y) \left[I_{\mathrm{L}}(y) \e^{\I (\theta-\varphi)}
    + I_{\mathrm{C}}(y) \e^{-\I \varphi}
    + (I_{\mathrm{R}}(y)-1) 
    \right]
    \:,
\end{multline}
\begin{multline}
    \label{eq:fctDblIntPtXY}
    f_t(X,Y,Y_0,\theta,\varphi) = [J_{\mathrm{C}}]^{K-1}
    \Bigg\lbrace
    G_t(X|0) G_t(Y|Y_0)
    + G_t(X|Y_0) G_t(Y|0)
    \\
   + \left(
        \e^{-\I \theta}\int_{-\infty}^0 \rho_0(y) G_t(Y|y) \dd y
        + \frac{\int_{0}^{Y_0} \rho_0(y) G_t(Y|y) \dd y}{J_{\mathrm{C}}}
        + \e^{-\I \varphi}\int_{Y_0}^\infty \rho_0(y) G_t(Y|y) \dd y
    \right)
    \\
    \times
    \Bigg[
    G_t(X|0) \left[I_{\mathrm{L}}(Y_0) \e^{\I \theta}
    + I_{\mathrm{C}}(Y_0)
    + I_{\mathrm{R}}(Y_0) \e^{\I \varphi}
    \right]
    + G_t(X|Y_0) \left[I_{\mathrm{L}}(0) \e^{\I \theta}
    + I_{\mathrm{C}}(0)
    + I_{\mathrm{R}}(0) \e^{\I \varphi}
    \right]
    \\
    + G_t(Y|0) \left[I_{\mathrm{L}}(Y_0) \e^{\I \theta}
    + I_{\mathrm{C}}(Y_0)
    + I_{\mathrm{R}}(Y_0) \e^{\I \varphi}
    \right]
    + G_t(Y|Y_0) \left[I_{\mathrm{L}}(0) \e^{\I \theta}
    + I_{\mathrm{C}}(0)
    + I_{\mathrm{R}}(0) \e^{\I \varphi}
    \right]
    \Bigg]
    \\
    + \left[I_{\mathrm{L}}(0) \e^{\I \theta}
    + I_{\mathrm{C}}(0)
    + I_{\mathrm{R}}(0) \e^{\I \varphi}
    \right]
    \left[I_{\mathrm{L}}(Y_0) \e^{\I \theta}
    + I_{\mathrm{C}}(Y_0)
    + I_{\mathrm{R}}(Y_0) \e^{\I \varphi}
    \right] \times \Bigg[
    \\
    \left(
        \e^{-\I \theta} \int_{-\infty}^0 \rho_0(y)G_t(Y|y) \dd y
        + \frac{\int_{0}^{Y_0} \rho_0(y) G_t(Y|y) \dd y}{J_{\mathrm{C}}}
        + \e^{-\I \varphi} \int_{Y_0}^\infty \rho_0(y) G_t(Y|y) \dd y
    \right)
    \\
    \times \left( \e^{-\I \theta}\int_{-\infty}^0 \rho_0(y)G_t(X|y) \dd y
    + \e^{-\I \varphi} \int_{Y_0}^\infty \rho_0(y)G_t(Y|y) \dd y \right)
    \\
    + \frac{\int_{0}^{Y_0}\rho_0(y)G_t(Y|y) \dd y}{J_{\mathrm{C}}}
        \left(
        \e^{-\I \theta}\int_{-\infty}^0 \rho_0(y)G_t(Y|y) \dd y
        + \frac{K-2}{K-1}\frac{\int_{0}^{Y_0} \rho_0(y)G_t(Y|y) \dd y}{J_{\mathrm{C}}}
        + \e^{-\I \varphi}\int_{Y_0}^\infty \rho_0(y)G_t(Y|y) \dd y
    \right)
    \Bigg]
    \Bigg\rbrace
    \:.
\end{multline}

Similarly, one can compute from~\eqref{eq:JointTrDens} the conditional profile, which takes the form
\begin{multline}
    \lim_{N \to \infty} \frac{ \moyI{\moyE{ \rho(x,t) \delta(x_0(t)-X) \delta(x_K(t)-Y)}} }{ \moyI{\moyE{ \delta(x_0(t)-X) \delta(x_K(t)-Y)}} }
    = a_t^{(i)}(X,Y|Y_0) G_t(x|0) + b_t^{(i)}(X,Y|Y_0) G_t(x|Y_0)
    \\
    + c_t^{(i)}(X,Y|Y_0) \int_{-\infty}^0 \rho_0(y) G_t(x|y) \dd y
    + d_t^{(i)}(X,Y|Y_0) \int_{0}^{Y_0} \rho_0(y) G_t(x|y) \dd y
    + e_t^{(i)}(X,Y|Y_0) \int_{Y_0}^\infty \rho_0(y) G_t(x|y) \dd y
    \:,
\end{multline}
where the superscripts $i=\mathrm{L},\mathrm{C}, \mathrm{R}$ respectively stand for $x < X$, $X<x<Y$, $Y<x$. The coefficients $a_t$, $b_t$, $c_t$, $d_t$ and $e_t$ are given by ratios of integrals of functions of the form~\eqref{eq:fctDblIntPtXY}, which are rather cumbersome, so we do not write them explicitly here. The important point is that these coefficients do not depend on $x$, so that the dependence of the conditional profile in $x$ is simply given by the propagator or its integral with the initial density.

\subsection{Long time limit: alternative derivation}

The joint distribution of the two tracers simplifies in the long time limit, if the propagator has, a scaling form
\begin{equation}
    \label{eq:ScalingProp}
    G_t(x|y) = \frac{1}{\sigma_t} g \left( \frac{x-y}{\sigma_t} \right)
    \:,
\end{equation}
with $\sigma_t \to \infty$ when $t \to \infty$. Let us also rescale the different parameters by $\sigma_t$,
\begin{equation}
    \label{eq:Scalings}
    X = \xi \: \sigma_t
    \:,
    \quad
    Y = \xi' \: \sigma_t
    \:,
    \quad
    Y_0 = z \: \sigma_t
    \:.
\end{equation}
One could use these scalings into~\eqref{eq:PtXY} and evaluate the integrals with a saddle point method. This is however quite tricky to do in practice. Instead, one could use an alternative approach, used in~\cite{Imamura:2017SM,Imamura:2021SM} with one tracer. Here, we extend this method to two tracers. The main idea is to consider the generalised current
\begin{equation}
    J_t(X|X_0) = \sum_{n} \left[ \Theta(x_i(t) - X) - \Theta(x_i(0) - X_0) \right] 
    \:
\end{equation}
This observable measures the difference between the number of particles at the right of $X$ at time $t$ and the number of particles to the right of $X_0$ at $t=0$. We define the tracer to be the first particle at the right of $X_0$ at $t=0$. Its position random, but this effect will be negligible in the long time limit because of the rescaling of space~\eqref{eq:Scalings}. Since the order of the particles is conserved, the position of the tracer at time $t$ can be found by finding $X$ such that $J_t(X|X_0) = 0$. The distribution of the position of the tracer is obtained, in the long time limit, by $P_t(X) = \mathbb{P}[J_t(X|X_0) = 0]$~\cite{Imamura:2017SM,Imamura:2021SM}.

In the case of two tracers, we need the joint distribution of two currents,
\begin{equation}
    P_t(X,Y|X_0,Y_0) = \mathbb{P}[J_t(X|X_0) = 0 \text{ and } J_t(Y|Y_0) = 0]
    \:.
\end{equation}
In the following, we will set $X_0 = 0$, and assume $Y_0>0$, and thus $Y > X$.

The first step is to consider the joint cumulant generating function of the two currents,
\begin{equation}
   \psi(\lambda,\mu) \equiv \lim_{N \to \infty} \ln \moyI{\moyE{ \e^{\lambda J_t(X|0) + \mu J_t(Y|Y_0)} }}
    = \int \dd y \: \rho_0(y) \int G_t(x|y) (\e^{\lambda (\Theta(x-X) - \Theta(y)) + \mu(\Theta(x-Y) - \Theta(y-Y_0)))}-1) \dd x
    \:,
\end{equation}
which follows directly from~\eqref{eq:PsiAnnealedGenSFSM} with $f(x) = \lambda \Theta(x-X) + \mu \Theta(x-Y)$ and $g(x) = \lambda \Theta(x) + \mu \Theta(x-Y_0)$. Using the scaling form~\eqref{eq:ScalingProp} for the propagator, with the definitions~\eqref{eq:Scalings}, we get, in the case of a constant density $\rho_0(x) = \rho$,
\begin{multline}
    \psi(\lambda,\mu) = \rho \: \sigma_t \left\lbrace 
        (\e^{-\lambda}-1) [h(-\xi) - h(z-\xi)] 
        + (\e^{-\lambda-\mu} - 1) h(z-\xi)
        + (\e^{\lambda}-1) [h(\xi) - h(\xi')]
    \right.
        \\
    \left.
        + (\e^{-\mu}-1) [h(z-\xi') - h(z-\xi)]
        + (\e^{\lambda+\mu}-1) h(\xi')
        + (\e^{\mu}-1) [h(\xi'-z) - h(\xi')]
    \right\rbrace
    \:,
\end{multline}
where we have introduced the double primitive of the propagator
\begin{equation}
    \label{eq:DblPrimProp}
    h(x) = \int_{x}^\infty \dd u \int_{u}^\infty \dd v \: g(v)
    \:.
\end{equation}
The joint distribution of the two currents can be obtained by a double Laplace inversion (in $\lambda$ and $\mu$) of $\e^{\sigma_t \psi(\lambda,\mu)}$. In the long time limit, the integration can be performed by a saddle point method, which reduces to a Legendre transform,
\begin{equation}
    \mathbb{P}\left[J_t(X|0) = \sigma_t q \text{ and } J_t(Y|Y_0)= \sigma_t q'\right]
    \simeq \e^{- \sigma_t \phi(q,q')}
    \:,
    \quad
    \phi(q,q') = -\psi(\lambda^\star,\mu^\star) + \lambda^\star q + \mu^\star q'
    \:,
\end{equation}
where $\lambda^\star$ and $\mu^\star$ are given by
\begin{equation}
    \left. \partial_\lambda \psi(\lambda,\mu) \right|_{\lambda^\star,\mu^\star} = q
    \:,
    \quad
    \left. \partial_\mu \psi(\lambda,\mu) \right|_{\lambda^\star,\mu^\star} = q'
    \:.
\end{equation}
The joint distribution of the two tracers is obtained by setting $q=q'=0$, and takes the large deviation form,
\begin{equation}
    P_t(X,Y|0,Y_0) \simeq \e^{-\sigma_t \phi(0,0)}
    \:.
\end{equation}
We can obtain the behaviour of the distribution near around the mean values $X = 0$ and $Y=Y_0$ by expanding $\lambda^\star$ and $\mu^\star$ in powers of $\xi$ and $\xi'-z$. This gives at leading order a quadratic behaviour for $\phi$,
\begin{equation}
    \phi(0,0) \simeq \frac{1}{2} \begin{pmatrix} \xi & \xi'-z  \end{pmatrix}
    \begin{pmatrix} 
    \frac{h(0)}{2h'(0)^2} & \frac{h(0)}{2h'(z)^2} \\
    \frac{h(0)}{2h'(z)^2} & \frac{h(0)}{2h'(0)^2}
    \end{pmatrix}^{-1}
    \begin{pmatrix} \xi \\ \xi'-z  \end{pmatrix}
    \:.
\end{equation}
This gives a Gaussian behaviour around the average values, with the fluctuations directly obtained from the covariance matrix,
\begin{equation}
    \boxed{
    \mathrm{Var}(X) = \mathrm{Var}(Y) =  \frac{h(0)}{2h'(0)^2} \sigma_t
    \:,
    \quad
    \mathrm{Cov}(X,Y) = \frac{h(z)}{2h'(0)^2} \sigma_t 
    \:.
    }
\end{equation}

\section{Conditional profiles vs correlation profiles}

We have introduced two equivalent ways to quantify the statistical properties of the position $x_0(t)$ of the tracer, and its correlations with the density $\rho(x,t)$ of particles:
\begin{itemize}
    \item[(i)] the cumulant generating function and the correlation profile
    \begin{equation}
        \psiA(\lambda,t) = \ln \moyI{ \moyE{\e^{\lambda x_0(t)}} }
        \:,
        \quad
        \wA(x,\lambda,t) = \frac{
        \moyI{ \moyE{ \rho(x,t) \e^{\lambda x_0(t)} } }
        }{
        \moyI{ \moyE{ \e^{\lambda x_0(t)} } }
        }
        \:,
    \end{equation}
    \item[(ii)] the distribution and the conditional profile
    \begin{equation}
        P_t(X) = \moyI{ \moyE{ \delta(X-x_0(t)) } }
        \:,
        \quad
        \tilde{w}_{\mathrm{A}}(x,X,t) = \frac{
        \moyI{ \moyE{ \rho(x,t) \delta(X-x_0(t)) } }
        }{
        \moyI{ \moyE{ \delta(X-x_0(t)) } }
        }
        \:.
    \end{equation}
\end{itemize}
The two formulations are related by Laplace transforms,
\begin{equation}
    \label{eq:RelProfLaplace}
    \e^{\psiA(\lambda,t)} = \int_{-\infty}^{\infty} \e^{\lambda X} P_t(X) \dd X
    \:,
    \quad
    \wA(x,\lambda,t) = \frac{
    \int_{-\infty}^\infty  \e^{\lambda X}\tilde{w}_{\mathrm{A}}(x,X,t) P_t(X) \dd X
    }{
    \int_{-\infty}^\infty  \e^{\lambda X} P_t(X) \dd X
    }
    \:.
\end{equation}
The correlation profiles can therefore be obtained from the conditional profile and vice versa. We will now see that in the long time limit, this relation greatly simplifies.

We start from the scaling form of the propagator~\eqref{eq:ScalingProp}, with a step initial density $\rho_0(x) = \rho_- \Theta(-x) + \rho_+\Theta(x)$. Plugging these into the distribution~\eqref{eq:DistrXtallTimes}, we obtain that the term $\phi_{X,t}(\theta)$ in the exponential is proportional to $\sigma_t$. The integral over $\theta$ can thus be evaluated in the limit $t \to \infty$ by a saddle point method. This gives that $P_t(X)$ takes a large deviation form,
\begin{equation}
    P_t(X) \underset{t \to \infty}{\simeq} \e^{- \sigma_t \chi( \xi = X/\sigma_t)}
    \:,
    \quad
    \chi(\xi) = \left( \sqrt{\rho_+ h(-\xi)} - \sqrt{\rho_- h(\xi)}  \right)^2
    \:,
\end{equation}
where $h$ is the double primitive of the propagator~\eqref{eq:DblPrimProp}. This generalises the formula given in~\cite{Sadhu:2015SM} for Brownian particles, and in~\cite{Hegde:2014SM} for a flat density $\rho_+ = \rho_-$. Importantly, this large deviation form allows to also evaluate the integrals in~\eqref{eq:RelProfLaplace} by a saddle point, which gives,
\begin{equation}
    \psiA(\lambda,t) \underset{t \to \infty}{\simeq} 
    \sigma_t \left[ \lambda \xi^\star(\lambda) - \chi(\xi^\star(\lambda)) \right]
    \:,
    \quad \text{and} \quad
    \wA(x,\lambda,t) \underset{t \to \infty}{\simeq}
    \tilde{w}_{\mathrm{A}}(x, \sigma_t \xi^\star(\lambda) ,t)
    \quad \text{where} \quad 
    \chi'(\xi^\star(\lambda)) = \lambda
    \:.
\end{equation}
In this limit, the conditional profile and the correlation profile are thus equal. But this is not the case at arbitrary time. It turns out that the correlation profiles are easier to compute for observables of the form~\eqref{eq:GenObsSM}, while the conditional profiles are more directly obtained for the position of a tracer.

\end{document}